\documentclass{article}
\usepackage{epsfig}
\usepackage{xspace}
\usepackage[latin1]{inputenc}
\usepackage{a4wide}
\def\S12{\mathrm{S}_{1,2}}

\newcommand{\xl}{{\em xloops}\xspace}
\newcommand{\halb}{\mbox{$\frac{1}{2}$}}

\newcounter{allequation}

\begin{document}

\title{\bf Experimental Signatures of Fermiophobic Higgs
  bosons\\[1cm]}
\author{L. Brücher${}^{1}$\\[0.5cm]
  {\small \em Centro de F\'\i sica Nuclear da Universidade de Lisboa,}\\
  {\small \em Av. Prof. Gama Pinto 2, 1699 Lisboa, Portugal}\\[0.5cm]
  {\small and}\\[0.5cm]
  R. Santos${}^2$\\[0.5cm]
  {\small \em CFNUL, Av. Prof. Gama Pinto 2, 1699 Lisboa, Portugal}\\
  {\small \em and}\\
  {\small \em Instituto Superior de Transportes, Campus Universit\'ario,}\\
  {\small \em R. D. Afonso Henriques, 2330 Entroncamento,
    Portugal}\\[1.5cm]} 
\date{20.07.1999}
\maketitle

\begin{abstract}
  The most general Two Higgs Doublet Model potential without explicit
  $CP$ violation depends on 10 real independent parameters. Excluding
  spontaneous $CP$ violation results in two 7 parameter models.
  Although both models give rise to 5 scalar particles and 2 mixing
  angles, the resulting phenomenology of the scalar sectors is
  different.
  
  If flavour changing neutral currents at tree level are to be
  avoided, one has, in both cases, four alternative ways of introducing the fermion
  couplings. In one of these models the mixing angle of
  the $CP$ even sector can be chosen in such a way that the fermion
  couplings to the lightest scalar Higgs boson vanishes. At the same time it is
  possible to suppress the fermion couplings to the charged and
  pseudo-scalar Higgs bosons by appropriately choosing the mixing angle of
  the $CP$ odd sector.
  
  We investigate the phenomenology of both models in the fermiophobic
  limit and present the different branching ratios for the decays of
  the scalar particles. We use the present experimental results from
  the LEP collider to constrain the models.
\end{abstract}
\vspace*{1cm}

\begin{flushleft}
  PACS number(s): 12.60.Fr, 14.80.Cp
\end{flushleft}

\footnotetext[1]{e-mail: bruecher@alf1.cii.fc.ul.pt}
\footnotetext[2]{e-mail: rsantos@alf1.cii.fc.ul.pt}

\thispagestyle{empty}



\newpage

\section{Introduction}

The $SU(2)\times U(1)$ electroweak model describes our world at the
presently attainable energies. Nevertheless, it is hard to hide the
frustration about our ignorance on the mass generation mechanism.
The spontaneous symmetry breaking mechanism requires a single doublet
of complex scalar fields. But does nature follow this minimal version
or does it require a multi-Higgs sector?

The current search at LEP already constrains the mass of a neutral
Higgs boson with a standard model like coupling to the fermions to
$m_H>91.0\ GeV$ \cite{SMH}. Nevertheless some multi-Higgs models
allow the existence of Higgs particles with a vanishing coupling to
the fermions. In this paper we investigate all type I $CP$ conserving Two
Higgs Doublets models (2HDM) with such a vanishing coupling to the
fermions. We will predict the experimental signatures of these
particles.

Our paper is organized as follows: first we review the different 2HDM
potentials to fix our notation.  Thereafter we try to restrict the
physical parameters of the potentials with theoretical constraints.
Then we will discuss the signature of the different particles and show
all characteristic branching ratios.  Finally we constrain the models'
parameters using the current experimental data.

\section{The potentials}

The minimal version of the standard model which allows spontaneous
symmetry breaking requires one scalar doublet of complex fields. To
assure the renormalizability of the theory, the most general potential
is
\begin{equation}
V=-\mu^{2} \phi^{\dagger} \phi + \lambda
\left( \phi^{\dagger} \phi \right)^2 \enskip ,
\end{equation}
where $\mu$ and $\lambda$ are real independent parameters. The mass
eigenstate is a $C$-even scalar particle, the Higgs boson.

The simplest generalization of the potential amounts to the
introduction of a second doublet of complex fields.  The most general
renormalizable potential invariant under $SU(2)\times U(1)$ has
fourteen independent real parameters. The number of predicted
particles grows from one to five. If one imposes that the potential is
invariant under charge conjugation $C$, the number of parameters is
reduced to ten.  Defining $x_1 = \phi_1^\dagger \phi_1$, $x_2 =
\phi_2^\dagger \phi_2$, $x_3 = \Re\{\phi_1^\dagger \phi_2\}$ and $x_4
= \Im\{\phi_1^\dagger \phi_2\}$ it can be shown \cite{Sant3} that the
most general 2HDM potential without explicit $C$ violation is:
\begin{equation}
V=-\mu_{1}^{2} x_{1} - \mu_{2}^{2}x_{2} - \mu_{12}^{2}x_{3}
 + \lambda_{1}x_{1}^2 + \lambda_{2}x_{2}^2 + \lambda_{3} x_{3}^2 
 + \lambda_{4}x_{4}^2 + \lambda_{5}x_{1}x_{2} + \lambda_6 x_1 x_3 
 + \lambda_7 x_2 x_3 \enskip , 
\end{equation}
where $\mu_i$ and $\lambda_i$ are real independent parameters.  The
number of parameters can be further reduced and there are two ways to
accomplish it. First, the potential can be made invariant under the
$Z_2$ transformation $\phi_1 \rightarrow \phi_1$ and $\phi_2
\rightarrow - \phi_2$. The resulting potential, which we call $V_A$,
is
\begin{equation}
V_{A}=-\mu_{1}^{2} x_{1}-\mu_{2}^{2} x_{2}+\lambda_{1} x_{1}^2
 +\lambda_{2} x_{2}^2+\lambda_{3} x_{3}^2+\lambda_{4} x_{4}^2 
 +\lambda_{5} x_{1} x_{2} \enskip . 
\end{equation}
If we allow a soft breaking term $-\mu_{12}^{2} x_{3}$ in $V_{A}$,
we end up with a model with spontaneous $CP$-violation \cite{bra1}.

Second, it is possible to make the potential invariant under the
global $U(1)$ transformation $\phi_2 \rightarrow e^{i \theta} \phi_2$.
The potential then reads:
\begin{equation}
V_{B}^\prime =-\mu_{1}^{2} x_{1}-\mu_{2}^{2} x_{2}
 +\lambda_{1} x_{1}^2+\lambda_{2} x_{2}^2
 +\lambda_{3} \left( x_{3}^2+ x_{4}^2 \right) 
 +\lambda_{5} x_{1} x_{2} \enskip . 
\end{equation}
Since we have a global broken symmetry, there is an extra Goldstone
boson in the theory. If we allow the same soft breaking term,
$-\mu_{12}^{2} x_{3}$, in the potential, we end up with the scalar
sector that has the same general structure as the scalar sector of the minimal
super symmetric model (MSSM) \cite{Gun}.\footnote{In the MSSM the
  $\lambda$´s are related to the gauge couplings $g$ and $g^\prime$.}  We
call this latter model $V_{B}$. Both $V_A$ and $V_B$ have seven
degrees of freedom, the five particle masses and the two rotation angles
($\alpha, \beta$).  The five particles can be grouped into 2 scalar
($h^0$,$H^0$), where the small letter denotes the less massive
particle, 1 pseudo-scalar particle ($A^0$) and 2 charged particles
($H^\pm$).\footnote{For the connection between the physical parameters
  and the original parameters from the potential see \cite{Sant4}.}
The main difference between the potentials is the self-couplings in
the scalar sector.

\section{The fermiophobic limit}

\begin{figure}[htbp]
  \begin{center}
    \epsfig{file=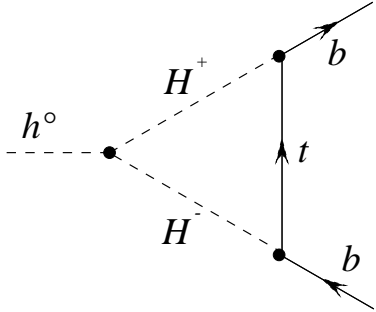,width=5cm}
    \caption{Feynman diagram of the largest contribution to $h^0\rightarrow b \bar{b}$}
    \label{fig:fgs}
  \end{center}
\end{figure}

Potentials $V_A$ and $V_B$ give rise to different self-couplings in
the scalar sector. However, the scalar couplings to the gauge bosons
and to the fermions are the same in both models. If flavour changing
neutral currents (FCNC) induced by Higgs exchanges are to be avoided,
one has four different ways to couple the scalars to the fermions. A
technically natural way to achieve it is to extend the global symmetry
to the Yukawa Lagrangian. This leads to two different ways of coupling
the quarks to the scalars as well as two different ways of coupling
the leptons to the scalars. The result is a total of four different
models, usually denoted as model I, II, III and IV (cf.  e.g.
\cite{Sant1}).

In model I, the lightest $CP$-even scalar, $h^0$, couples to a fermion
pair (quark or lepton) proportionally to $\cos\alpha$.  Setting
$\alpha = \pi/2$, $h^0$ becomes completely fermiophobic.  However,
$h^0$ can still decay to two fermion pair via \mbox{$h^0 \rightarrow
  W^{*}W (Z^{*}Z) \rightarrow 2\, \bar{f} f$} or \mbox{$h^0
  \rightarrow W^* W^* (Z^* Z^*) \rightarrow 2\,\bar{f} f$}. We will
include these decays in our analysis. It is worthwhile to point out
that these processes occur near the $W (Z)$ threshold.  Decays of
$h^0$ to two fermions can also be induced by scalar and gauge boson
loops (see e.g. fig.~\ref{fig:fgs}). In the 2HDM, the angle $\alpha$
has to be renormalized to render $h^0 \rightarrow f\bar{f}$ finite.
However, at $\alpha=\pi/2$, all one-loop decays $h^0 \rightarrow
f\bar{f}$ are finite. Thus we can impose the following condition for
$\delta\alpha$: the renormalized one-loop decay width for $h^0
\rightarrow f\bar{f}$ is equal to the finite unrenormalized decay
width. This condition is equivalent to set
$[\delta\alpha]_{\alpha=\pi/2}=0$. We have checked that this condition
holds for all fermions. The only relevant one-loop decay is $h^0
\rightarrow b\bar{b}$ due to a large contribution of the Feynman
diagram shown in fig.~\ref{fig:fgs} to the total decay
width.\footnote{The coupling $[H^+\bar{t}b]$ is proportional to the
  $t$-quark mass.}  Thus, on one hand, $h^0$ is not completely
fermiophobic at $\alpha=\pi/2$, and on the other hand, all decays $h^0
\rightarrow f\bar{f}$ but $h^0 \rightarrow b\bar{b}$ are almost zero
even at one-loop level.

The couplings of the $CP$-odd scalar, $A^0$, and of the charged
scalar, $H^{\pm}$, are proportional to $\cot \beta$. If we want these
particles to be fermiophobic as well, $\beta$ has to approach $\alpha$
($\beta\rightarrow\alpha =\pi/2$). In this limit the coupling of $h^0$
to the vector bosons, which is proportional to the sine of
$\delta\equiv \alpha -\beta$ tends to zero. Thus, $h^0$ is not only
fermiophobic but also bosophobic and ``ghostphobic'' -- $h^0$ always
needs another scalar particle to be able to decay. The differences
between potential $A$ and $B$ can be extremely important in this limit
since $h^0$ will have different signatures in each model. In contrast,
the heaviest $CP$-even scalar, $H^0$, acquires the Higgs standard
model couplings to the fermions in this limit. We will relax the limit
$\beta \approx \pi/2$ and analyze the decays as a function of $\delta$
and of the Higgs masses.

\section{Theoretical mass limits}\label{masslim}

\begin{figure}[htbp]
  \begin{center}
    \epsfig{file=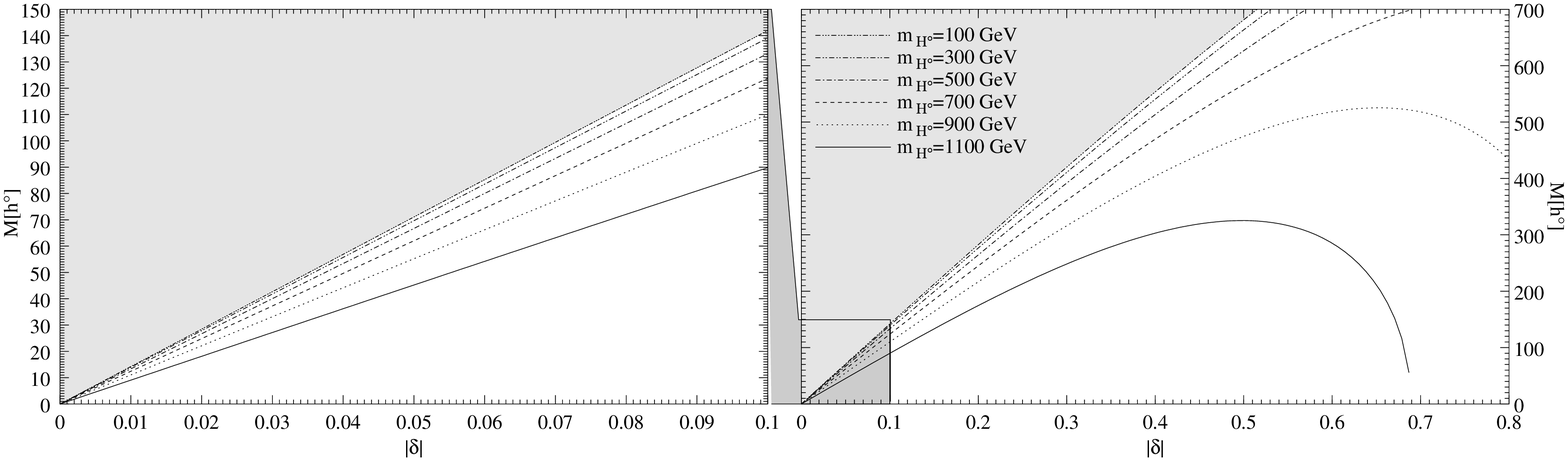,width=16.5cm,height=8.5cm}
    \caption{Limit on $m_{h^0}$ as a function of $\delta$ in potential$A$.}
    \label{fig:Mhlimit1A}
  \end{center}
\end{figure}

Although the parameters of the 2HDM's are, in contrast to the MSSM,
almost unconstrained, it is possible to derive some bounds on the
masses of the scalar sector particles in the fermiophobic limit. We want to look
for the allowed region in the $m_{h^0}$-$\beta$ plane so that the
calculations do not leave the perturbative regime. Several methods of
achieving theoretical bounds on these masses have been published
\cite{SB1}. Tree-level unitarity bounds have been derived in
\cite{trunal1} and \cite{trunal2} for potential $A$. We use the bounds
from \cite{trunal1}:
\begin{equation}
  \label{eq:vacA}
  m_{h^0} \, \le \, \sqrt{\frac{16\pi\sqrt{2}}{3\,G_F}\,\cos^2\beta \,
  - \, m_{H^0}^2 \cot^2\beta} \enskip ,
\end{equation}
where $G_F=1.166\ GeV^{-2}$ denotes Fermi´s constant. Equation
(\ref{eq:vacA}) is plotted in fig.~\ref{fig:Mhlimit1A}, where $\delta
= \pi/2 -\beta$ has been chosen for convenience.
Fig.~\ref{fig:Mhlimit1A} shows that in the limit $\delta\rightarrow 0$
the dependence on the angle is strong, whereas the dependence on
$m_{H^0}$ is very mild. In this limit $h^0$ is massless, which is
already clear from the definition of $m_{h^0}$ in the fermiophobic
limit \cite{Sant4}:
\begin{equation}
  m_{h^0}  =  \sqrt{2 \lambda_1 } \, v \, \cos \beta
\end{equation}

\begin{figure}[htbp]
  \begin{center}
    \epsfig{file=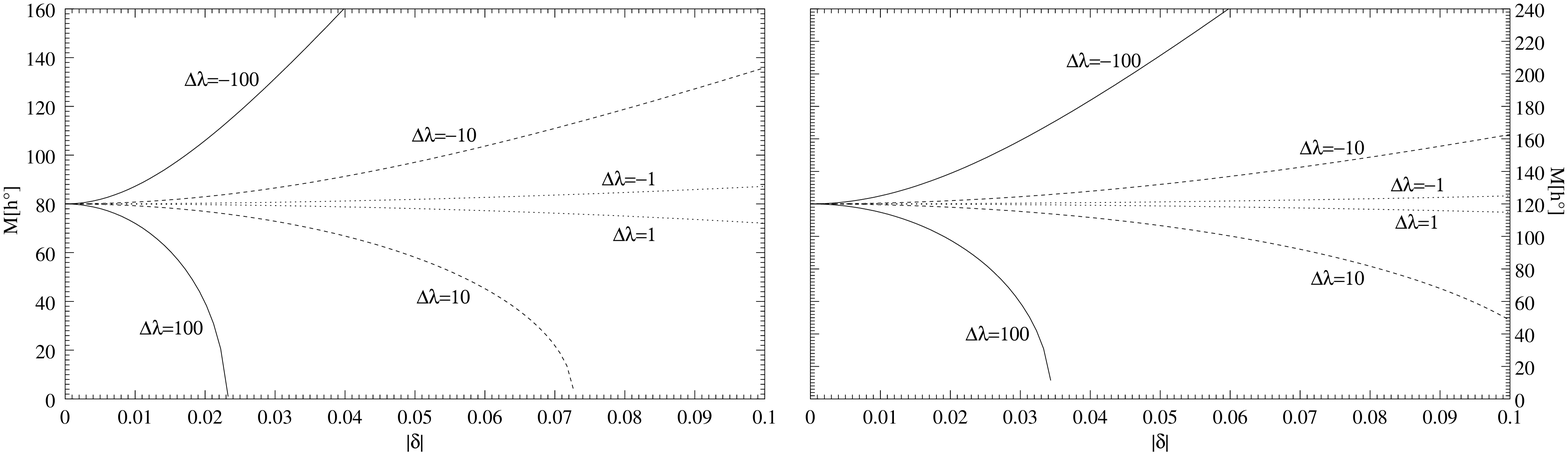,width=16cm,height=8.5cm}
    \caption{Limit $m_{h^0}$ as a function of $\delta$ for
      $m_A=80\ GeV$ and $m_A=120\ GeV$.}
    \label{fig:Mhlimit1B}
  \end{center}
\end{figure}

No tree-level unitarity bounds have been derived for potential $B$. A
full derivation of these limits would be beyond the scope of this
article. Nevertheless, we know that in the fermiophobic limit
\cite{Sant4}:
\begin{equation}
  \label{eq:vacB}
   m_{h^0}^2  =   m_A^2 - 2\left(\lambda_+ - \lambda_1\right) v^2
   \cos^2 \beta \enskip ,
\end{equation}
with $\lambda_+ = \halb(\lambda_3+\lambda_5)$ and $v=246\ GeV/c^2$ denoting the vacuum expectation value. The
equation shows, that in the limit $\delta\rightarrow 0$ the masses of
$h^0$ and $A^0$ will be degenerated. As already stated, stringent
bounds on all $\lambda{_i}$´s are missing.\footnote{However
  $\lambda_+=\frac{m_A^2}{2v^2}$, which means $0<\lambda_+<8.3$, if
  $m_A<1\ TeV$.}
On the other hand it might be sufficient to explore equation
(\ref{eq:vacB}) for different values of $ \lambda_+ - \lambda_1\equiv
\Delta\lambda$.
Fig.~\ref{fig:Mhlimit1B} shows $m_{h^0}$ as a function of $\delta$ for
different $\Delta\lambda$'s. On the left plot we set $m_A=80\ GeV$ and
on the right plot $m_A=120\ GeV$. The region limited by each value of
$|\Delta\lambda|$ is the allowed region for $m_{h^0}$ for the given
value of $m_A$. Although it is most likely that $|\Delta\lambda|<10$,
a value of $|\Delta\lambda|=100$ cannot strictly be excluded, if one
wants to be very conservative.



In potential $B$, if $\delta \le 0.05$, the masses of the lightest
scalar and the pseudo-scalar are almost degenerated.  $m_{h^0}$ can
differ at most 15\% from $m_{A}$, if $|\Delta\lambda|<10$.  In the
region where $0.05 < |\delta| \le 0.1$ the situation smoothly changes
for the same $\Delta\lambda$.  The restriction on the mass splitting
vanishes totally when $\delta \gg 0.1$.

In potential $A$ we have an upper limit for $m_{h^0}$ independent of
$m_A$. This upper bound depends on the value of $m_{H^0}$. It can be
as low as $45\ GeV$ or reach a maximum of approximately $m_W$, if
$\delta=0.05$. For $\delta=0.1$ the bound on $m_{h^0}$ varies between
$90\ GeV$ and $140\ GeV$. Very stringent bounds
on $m_{h^0}$ can be found, if $m_{H^0}$ is around $1\ TeV$ and
$\delta$ is large. On the
other hand for $m_{H^0}<700\ GeV$ no significant bounds can be found
for a wide range of $\delta$ values.

We want to stress again (see \cite{Sant4}) that respecting these
bounds is important to make reliable predictions on the decays of the
Higgs particles of the 2HDM. Otherwise, one runs into spurious
infinities of the couplings, which are not present in the original
parameters of the potential.

These theoretical bounds and the overall picture given by the
branching ratios shown in the next sections, led us to
distinguish between three different regions for $\delta$. For our
later qualitative analysis it is convenient to define the following
regions:
\begin{itemize}
\item the {\em tiny} $\delta$ region where $|\delta| \le 0.05$,
\item the {\em small} $\delta$ region with $0.05 < |\delta| \le 0.1$
  and
\item finally the medium and {\em large} $\delta$ region when $|\delta| > 0.1$.
\end{itemize}

\section{The lightest scalar Higgs boson}\label{h0}

As already pointed out, the lightest scalar Higgs boson ($h^0$) has no
tree level couplings to the fermions for $\alpha=\pi/2$. Thus the
following tree level decays have to be considered:
\begin{displaymath}
  h^0 \rightarrow W^+ W^- \quad ; \quad  
  h^0 \rightarrow Z Z \quad ; \quad 
  h^0 \rightarrow Z A^0 \quad ; \quad 
  h^0 \rightarrow W^\pm H^\mp \quad ; \quad 
  h^0 \rightarrow A^0 A^0 \quad ; \quad  
  h^0 \rightarrow H^+ H^- \quad .
\end{displaymath}
Additionally the following one-loop induced decays are important:
\begin{displaymath}
  h^0 \rightarrow \gamma\gamma \quad ; \quad 
  h^0 \rightarrow Z \gamma \quad ; \quad 
  h^0 \rightarrow b \bar{b} \quad .
\end{displaymath}
Moreover, decays to fermions via virtual vector bosons have to be
taken into account, namely:
\begin{displaymath}
   h^0 \rightarrow W^* W^* \rightarrow f\bar{f} f\bar{f} \quad ; \quad
   h^0 \rightarrow W^* W \rightarrow f\bar{f} W \quad ; \quad 
   h^0 \rightarrow Z^* Z^* \rightarrow f\bar{f} f\bar{f} \quad ; \quad
   h^0 \rightarrow Z^* Z \rightarrow f\bar{f} Z  \quad .
\end{displaymath}
The partial tree-level decay widths are listed in appendix \ref{Adw}.
The one-loop induced decays have been calculated with \xl \cite{xl1}.
Decays via virtual particles have been calculated in ref.~\cite{andr}. We
have taken these formulas and changed them appropriately. The decays
into one vector boson and one scalar have been calculated in this
paper for on-shell particles only. Near the thresholds decays via
virtual particles (i.e. $h^0 \rightarrow W^* H^\pm$ and $h^0
\rightarrow Z^* A^0$) can be taken into account. These decays have
been calculated in \cite{Akr}, where also formulas are given. The same
applies to all other scalar particle decays calculated in the
following sections.  

As stated earlier, the only significant decay mode to fermions, via
vector boson and scalar loops, is $h^0 \rightarrow b\bar{b}$. For all
the 
other fermionic decays the Feynman graphs are suppressed either by the
Cabbibo-Kobayashi-Maskawa matrix or by the small mass of the fermions
in the loop. However, the diagram shown in
fig.~\ref{fig:fgs} is suppressed by a $\tan^2 \delta$ factor when
compared with the corresponding diagram in $h^0 \rightarrow
\gamma\gamma$.  Thus, as will be seen below, the decay $h^0\rightarrow
b\bar{b}$ is of minor importance in the tiny and small $\delta$
region.

\begin{figure}[htbp]
  \begin{center}
    \epsfig{file=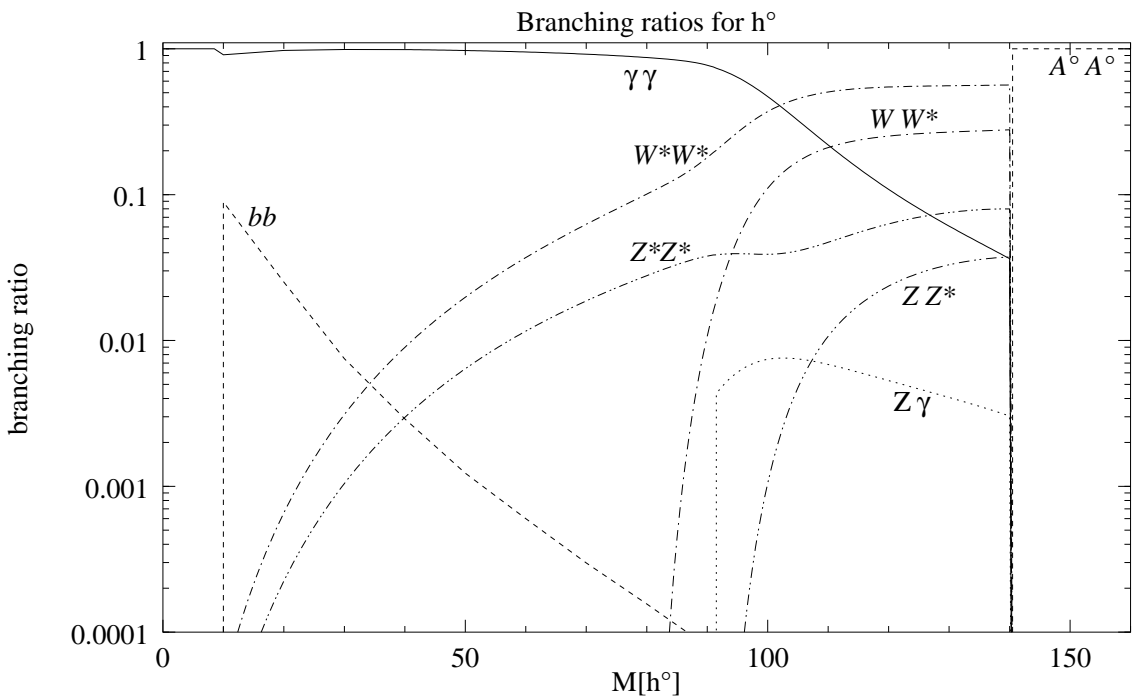,width=15cm}
    \caption{Branching ratios of $ h^0 $ at $m_{A^0}=70\ GeV$, $m_{H^+}=140\ GeV$, $m_{H^0}=300\ GeV$ and $\delta= 0.1$ in potential $A$. }
    \label{fig:h0A1}
  \end{center}
\end{figure}

\begin{figure}[htbp]
  \begin{center}
    \epsfig{file=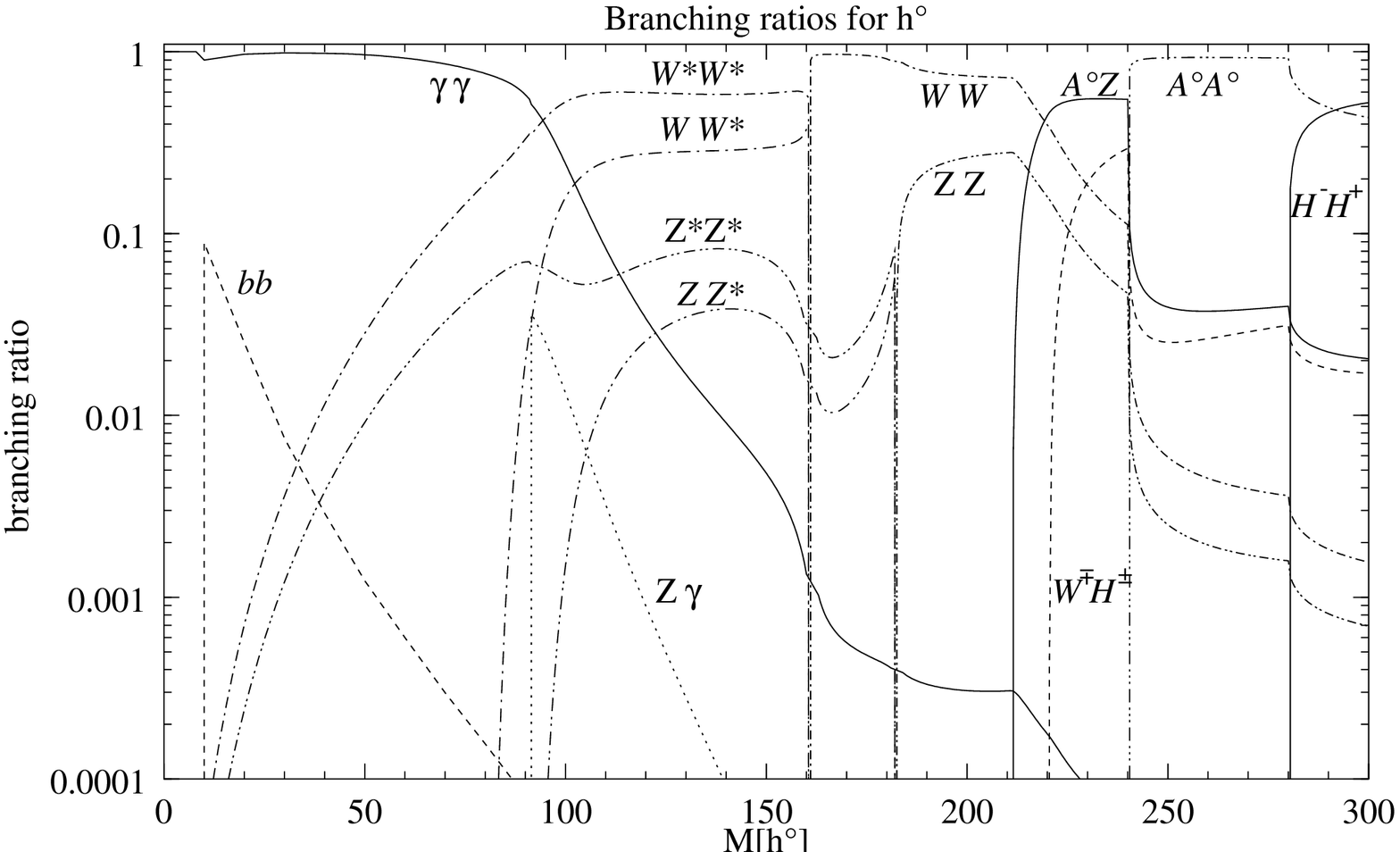,width=15cm}
    \caption{Branching ratios of $h^0$ at $m_{A^0}=120\ GeV$, $m_{H^+}=140\ GeV$, $m_{H^0}=300\ GeV$ and $\delta= 0.2$ in potential $A$. }
    \label{fig:h0A3}
  \end{center}
\end{figure}

In potential $A$ the upper bound for the mass of the lightest scalar
Higgs boson is approximately the $W$ mass in the tiny $\delta$ region.
Thus $h^0$ has only two possible decay modes. Either it decays into
$A^0 A^0$, if the mass of the lightest scalar is twice as large as the
mass of the pseudo-scalar Higgs boson, or it decays into two
photons.\footnote{The third possible decay, $h^0\rightarrow H^+H^-$ is
  already ruled out by the experimental lower limit on the mass of the
  charged Higgs boson (cf. section \ref{exlim}).} In the small
$\delta$ region the growth of the upper mass limit for $m_{h^0}$ gives
rise to more decay modes, as can be seen in fig.~\ref{fig:h0A1}. For
small $h^0$ masses the situation is the same as in the tiny $\delta$
region.  Depending on the mass of the pseudo-scalar, the dominant
decay is again either $h^0 \rightarrow A^0 A^0$ or
$h^0\rightarrow\gamma\gamma$. As soon as $m_{h^0} > m_W$, decays via
virtual vector bosons overtake the decay to $\gamma\gamma$ and give
rise to a fermionic signature of $h^0$.  Of course the value of
$m_{h^0}$, for which the branching ratio of $h^0\rightarrow W^*W^*$
becomes bigger than 50\% depends on $\delta$.  At the lower end of the
small $\delta$ region this happens approximately at $m_{h^0}=110\ 
GeV$, whereas at the upper end it is close to the $W$ mass. At first,
in the large $\delta$ region the branching ratio does not change much.
Of course the upper bound for $m_{h^0}$ looses importance and all
decays become kinematically allowed, as can be seen in
fig.~\ref{fig:h0A3}.  As $\delta$ increases, the decay $h^0\rightarrow
b \bar{b}$ becomes more and more significant for small masses of
$m_{h^0}$. If e.g. $m_{h^0}=20\ GeV$ we get a branching ratio for
$h^0\rightarrow b \bar{b}$ of the order of $30\%$ at $\delta=0.5$ and of
$75\%$ at $\delta=1.0$. This reflects the already mentioned
$\tan^2\delta$ suppression of this decay mode.


\begin{figure}[htbp]
  \begin{center}
    \epsfig{file=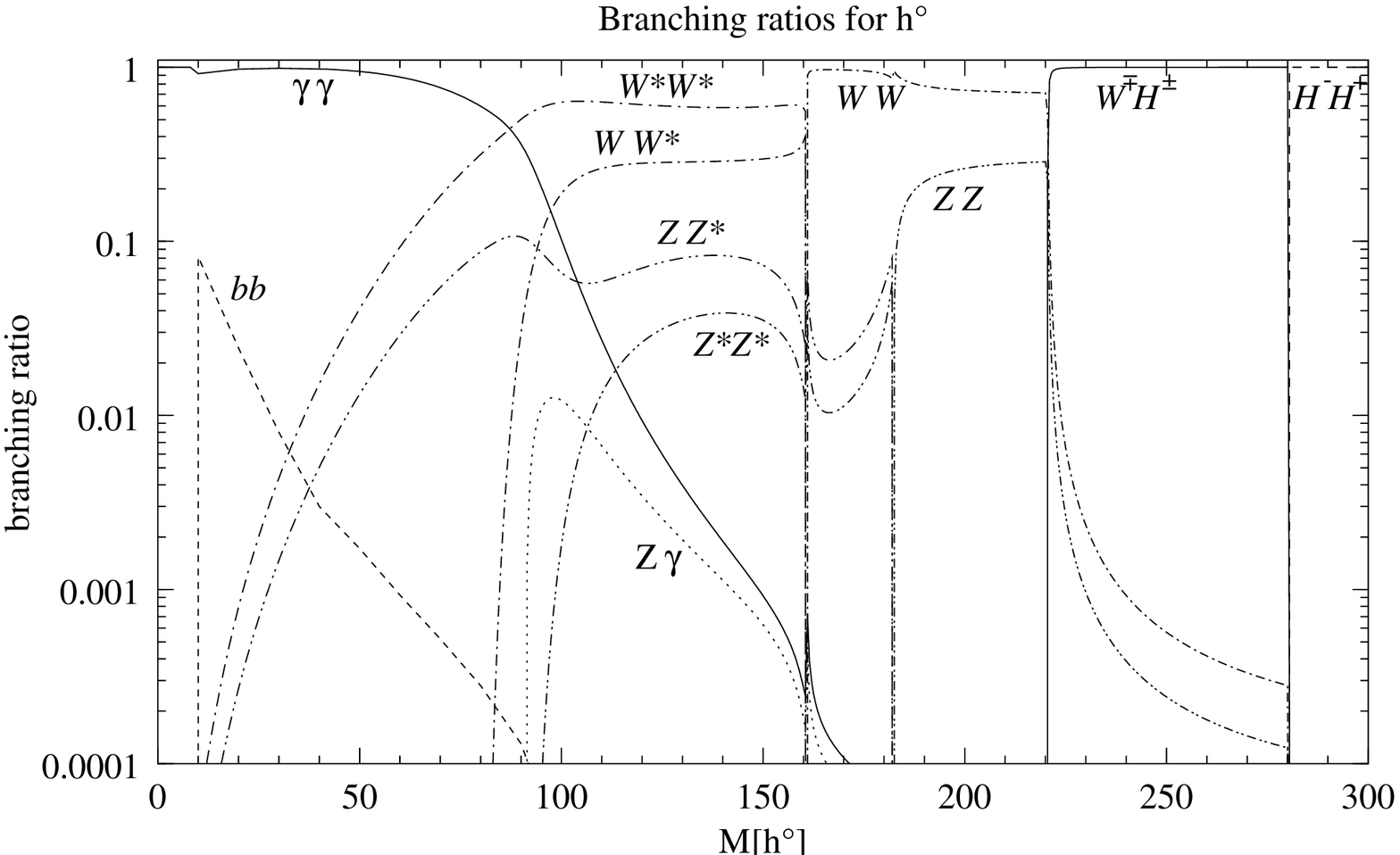,width=15cm}
    \caption{Branching ratios of $h^0$ at $m_{A^0}=m_{h^0}$, $m_{H^+}=140\ GeV$, $m_{H^0}=300\ GeV$ and $\delta= 0.01$ in potential $B$. }
    \label{fig:h0B1}
  \end{center}
\end{figure}

\begin{figure}[htbp]
  \begin{center}
    \epsfig{file=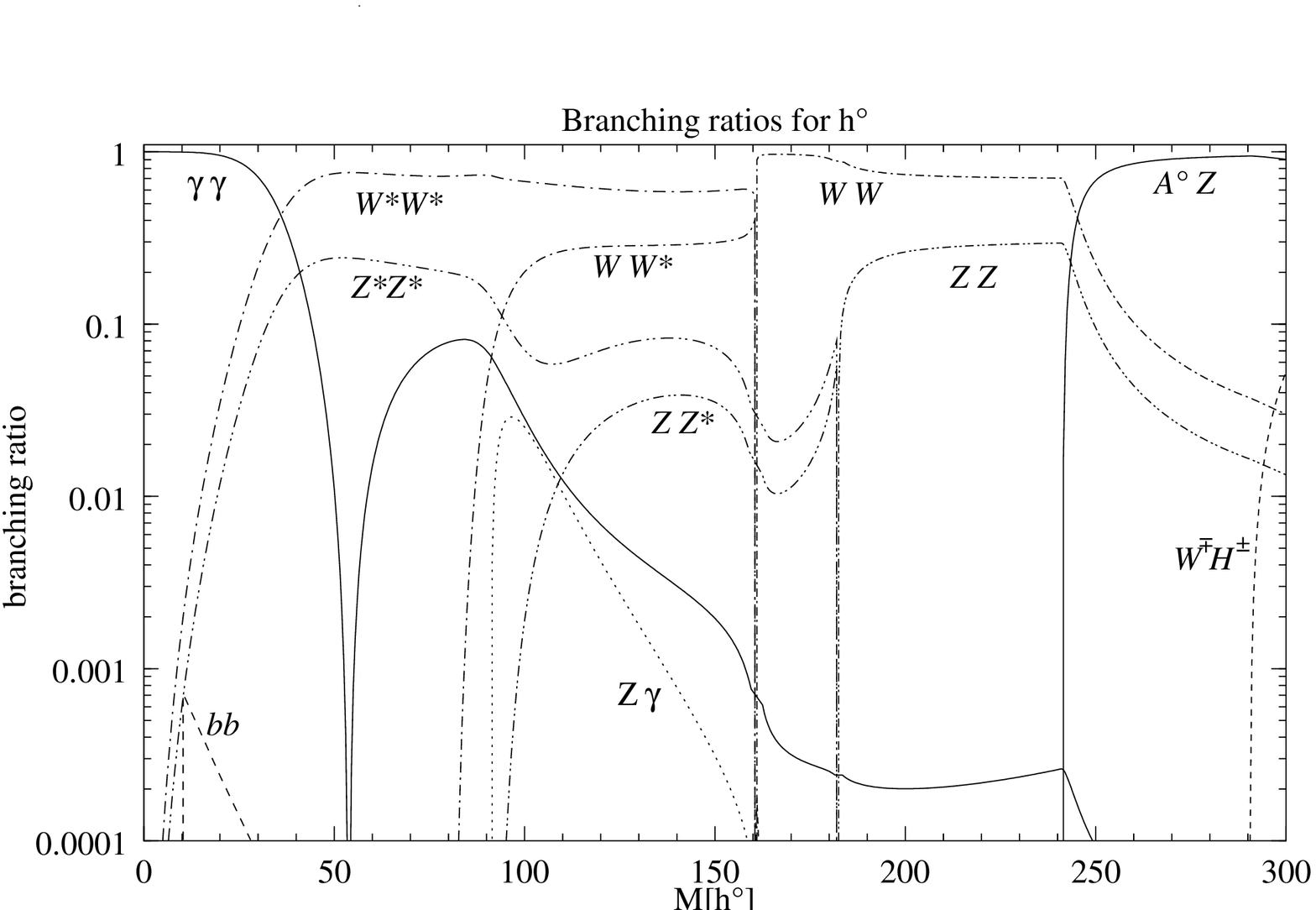,width=15cm}
    \caption{Branching ratios of $h^0$ at $m_{A^0}=150\ GeV$, $m_{H^+}=210\ GeV$, $m_{H^0}=300\ GeV$ and $\delta= 0.1$ in potential $B$. }
    \label{fig:h0B2}
  \end{center}
\end{figure}

In potential $B$ the masses of $h^0$ and $A^0$ are almost degenerated
in the tiny $\delta$ region. Thus for small masses ($<m_W$) $h^0$
decays mainly into two photons. On the other hand, no upper bound on
$m_{h^0}$ exists in potential $B$. As a consequence a heavy $h^0$ can
also decay via virtual vector bosons into fermions in the tiny
$\delta$ region (cf. fig.~\ref{fig:h0B1}).  In the small $\delta$
region the branching ratio strongly depends on the parameters $m_A$
and $m_{H^+}$. It can either resemble the plot for potential $A$ (see
fig.~\ref{fig:h0A1}), or, due to strong cancellation between the
$H^+$- and the $W$-loops in the $h^0\rightarrow\gamma\gamma$ decay, it
can be as shown in fig.~\ref{fig:h0B2}. In this figure we see that
$h^0\rightarrow\gamma\gamma$ only dominates until $m_{h^0}\approx 30\ 
GeV$. Then, decays via virtual vector bosons are the major decays of
$h^0$.  Note that $h^0\rightarrow b \bar{b}$ is suppressed in a
similar way to $h^0\rightarrow\gamma\gamma$, because both decays
depend on the same couplings of $h^0$ to the vector bosons and to the
scalars. In the large $\delta$ region this behaviour is almost the
same.  Of course, as in potential $A$, for some value of $\delta$ the
decay $h^0\rightarrow b \bar{b}$ will dominate over
$h^0\rightarrow\gamma\gamma$ for small values of $m_{h^0}$.

\begin{figure}[htbp]
  \begin{center}
    \epsfig{file=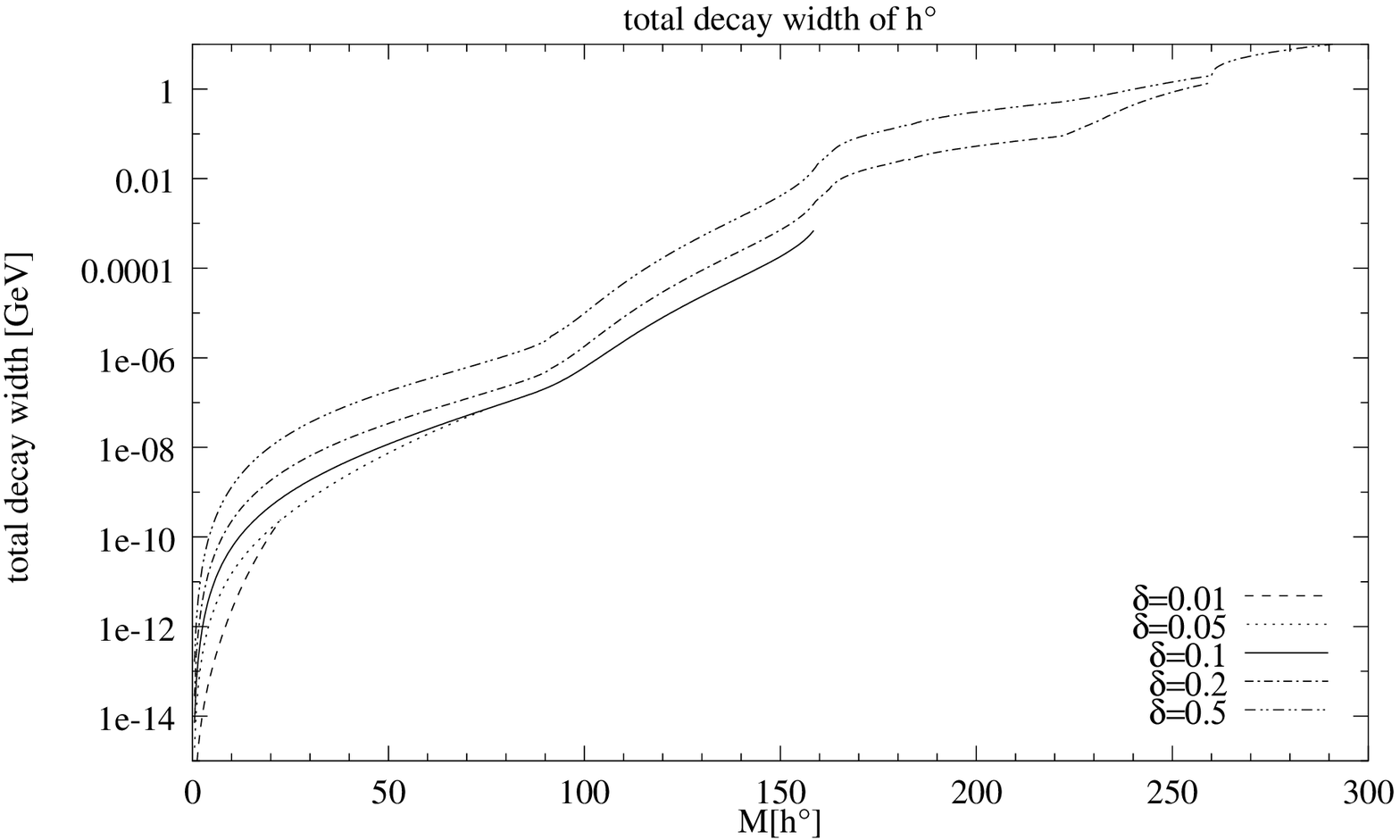,width=15cm}
    \caption{Total decay width of $h^0$ with $m_{A^0}=130\ GeV$,
      $m_{H^+}=150\ GeV$, $m_{H^0}=300\ GeV$ for different values of
      $\delta$ potential $A$. }
    \label{fig:h0A2}
  \end{center}
\end{figure}

Finally we show the total branching ratio of $h^0$ as function of
$m_{h^0}$ for different values of $\delta$ in fig.~\ref{fig:h0A2}. As
expected, the total decay width grows with $m_{h^0}$ and $\delta$.  We
do not show the total decay width for potential $B$ because the
overall behaviour is the same as for potential $A$.

\section{The pseudo-scalar Higgs boson}

For our analysis the following tree level decays have to be
considered:
\begin{displaymath}
  A^0 \rightarrow f\bar{f} \quad ; \quad A^0 \rightarrow Z h^0
  \quad ; \quad A^0 \rightarrow W^\pm H^\mp
\end{displaymath}
Furthermore the following one-loop decays have been calculated:
\begin{displaymath}
  A^0 \rightarrow \gamma\gamma \quad ; \quad 
  A^0 \rightarrow Z \gamma \quad ; \quad 
  A^0 \rightarrow W^+ W^- \quad ; \quad 
  A^0 \rightarrow ZZ \quad ; \quad 
  A^0 \rightarrow gg \quad ,
\end{displaymath}
where $g$ denotes a gluon.

 
Leaving aside the tree level decays into a final state with at least
one Higgs particles, all decays depend on the coupling of $A^0$ to the
fermions ($\cot^2 \beta$ in model I). This is a consequence of the
fact that without fermions $CP$ conservation is equivalent to separate
$C$ and $P$ conservation. So, no on-shell decay with only vector
bosons in the final state is possible. When the fermions are included,
$C$ and $P$ are no longer independently conserved, and so $A^0$ can
decay into two photons, for instance, via a fermion loop.


\begin{figure}[htbp]
  \begin{center}
    \epsfig{file=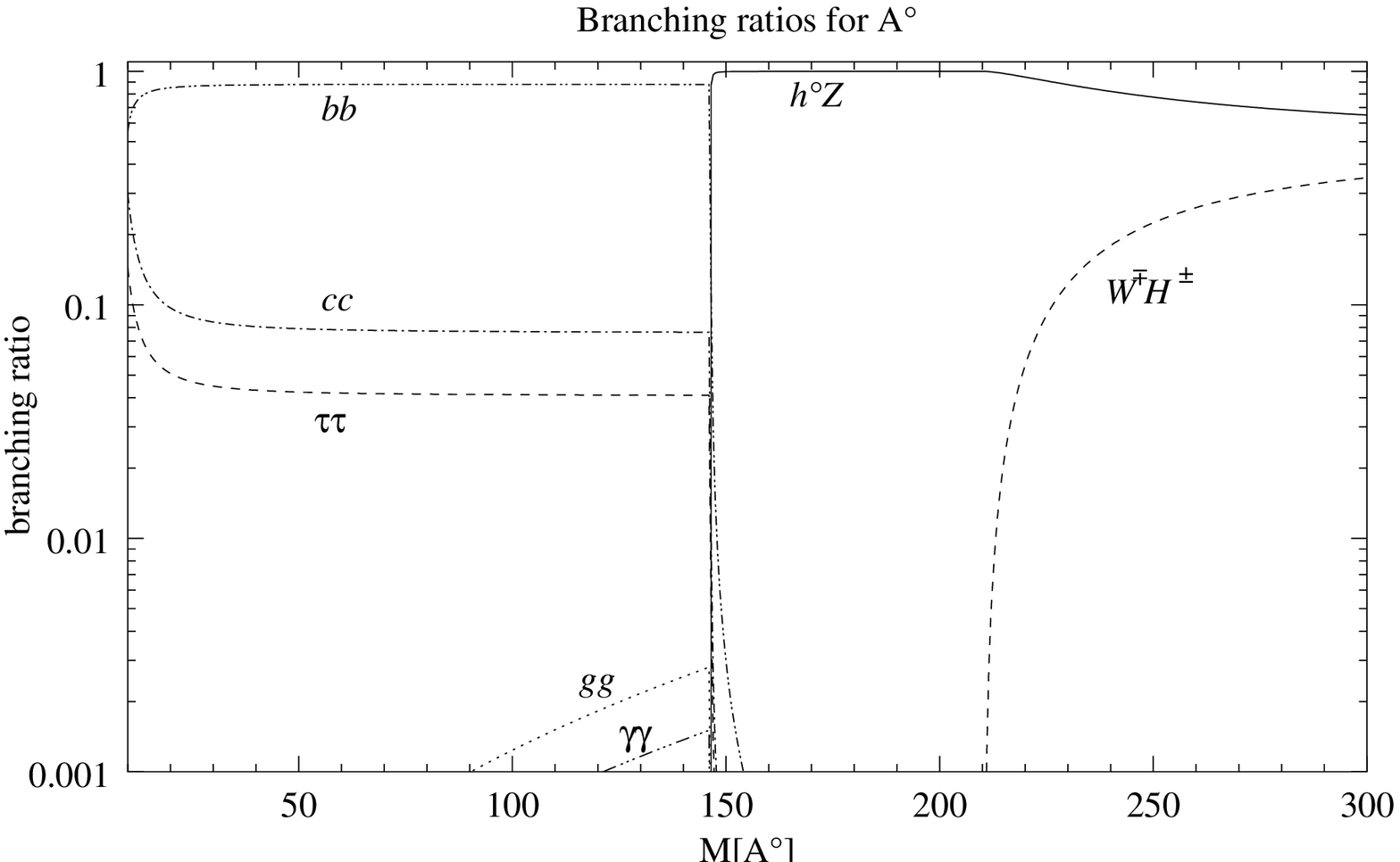,width=15cm}
    \caption{Branching ratios of $A^0$ as a function of the mass for $\delta=0.1$,
      $m_{h^0}=55\ GeV$, $m_{H^0}=300\ GeV$ and $m_{H^+}=130\ GeV$.}
    \label{fig:Abranch1}
  \end{center}
\end{figure}

\begin{figure}[htbp]
  \begin{center}
    \epsfig{file=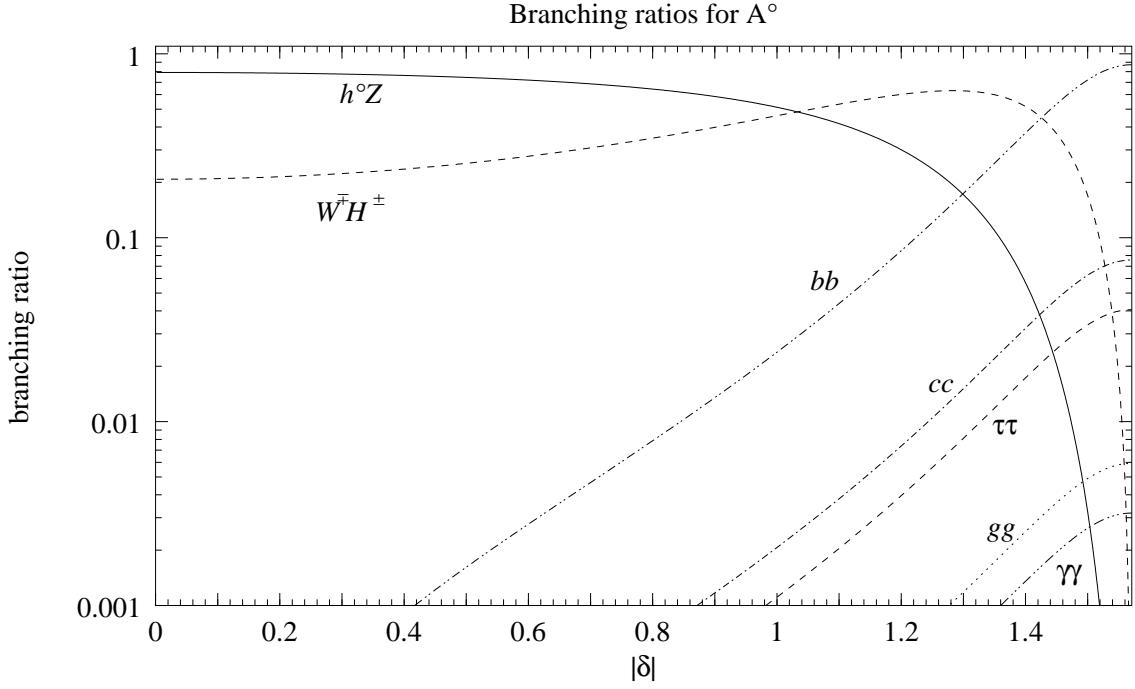,width=15cm}
    \caption{Branching ratios as a function of $\delta$ for $m_{A^0}=190\ GeV$, 
      $m_{h^0}=55\ GeV$, $m_{H^0}=300\ GeV$ and $m_{H^+}=100\ GeV$.}
    \label{fig:Abranch3}
  \end{center}
\end{figure}

\begin{figure}[htbp]
  \begin{center}
    \epsfig{file=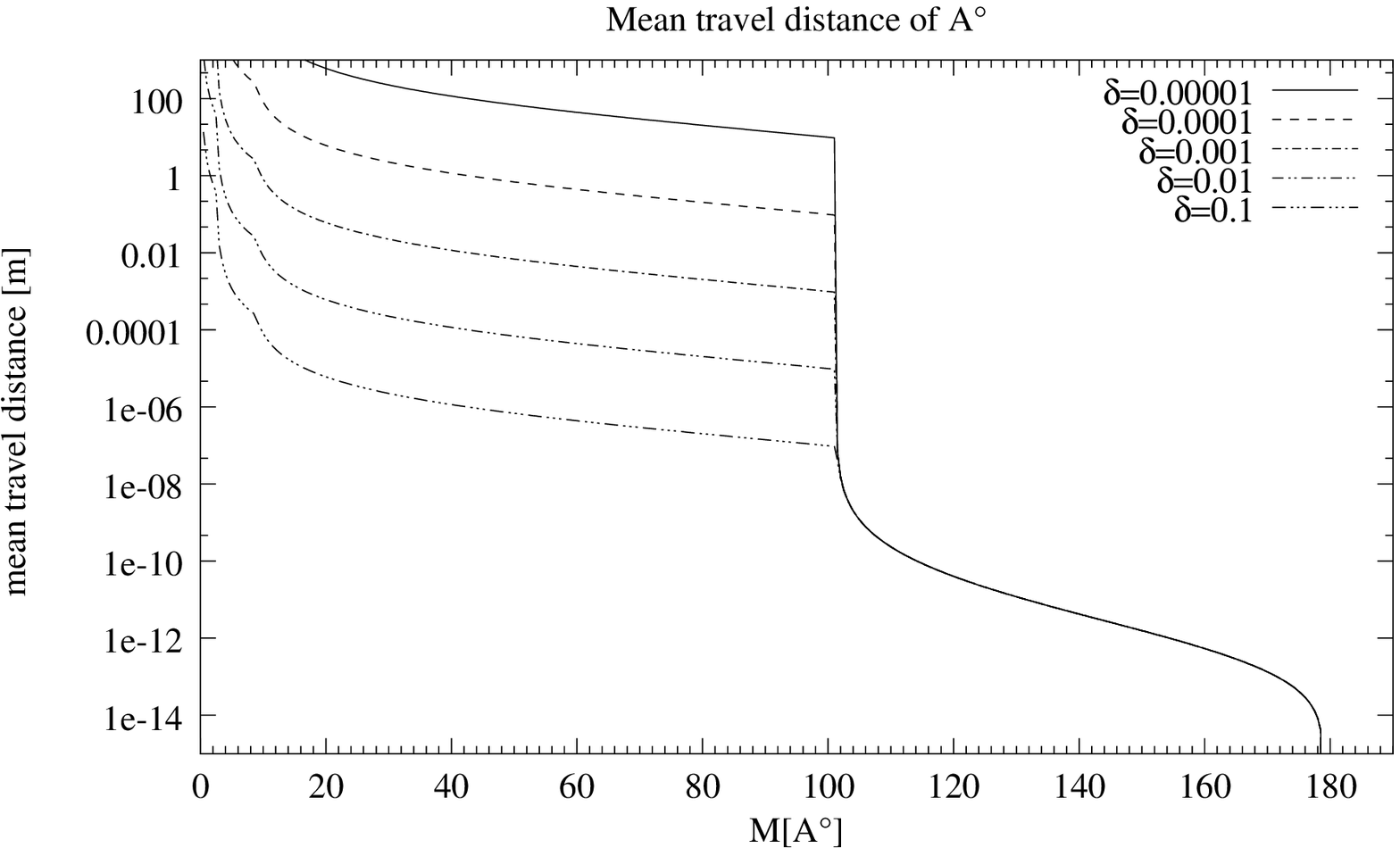,width=15cm}
    \caption{Travel distance of $A^0$ at $\sqrt{s}=189\ GeV$ and $m_{h^0}=10\ GeV$.}
    \label{fig:Abranch4}
  \end{center}
\end{figure}

On the other hand, when fermions are added, $A^0$ may directly decay
into them. As these are tree-level decays, their partial decay widths
are obviously larger than one-loop induced decay widths. This can also
be seen in fig.~\ref{fig:Abranch1}. This figure clearly shows, that
the one-loop decays of the pseudo-scalar Higgs are in the per mille
region when compared to the fermionic decays. We have checked that 
the $A^0$ branching ratio is independent of $\delta$ below $Z h^0$
and $W^\pm H^\mp$ thresholds. As we have pointed out, in this region all
decays just depend on $\cot \beta$.  This dependence cancels in the
branching ratios, but not in the decay width. Above these thresholds
$A^0$ decays mainly into $Z h^0$ and $W^\pm H^\mp$, as can be seen in
fig.~\ref{fig:Abranch3}. Only in the very large $\delta$ region the
decays into fermions dominate due to the dependence on $\tan \delta$.

Although below the $Z h^0$ and the $W^\pm H^\mp$ thresholds $A^0$ decays
mainly into fermions, the total decay width of $A^0$ decreases with
$\tan^2\delta$ in this region. So in the limit $\delta\rightarrow 0$
the pseudo-scalar Higgs will be a stable particle leaving no
characteristic signature in the detector. Furthermore, for a
sufficiently small $\delta$, $A^0$ decays outside the detector (see
fig.~\ref{fig:Abranch4}). 
So, the only way to detect it in this $\delta$ region is to consider
reactions with missing energy and momentum in the final state. The
situation changes, as soon as the $Z h^0$ or the $W^\pm H^\mp$
thresholds are crossed. Then $A^0$ decays inside the detector with
either a $Z h^0$ or a $W^\pm H^\mp$ signature.

Finally we notice, that all decays either depend on the coupling of
$A^0$ to fermions or to vector bosons. There is no decay, where
couplings of the scalars among themselves contribute to the decay
width. Thus for the pseudo-scalar Higgs boson no difference between
potential $A$ and $B$ can be seen in the branching ratios and decay
widths. So, the signature of the $A^0$ may be called Higgs potential
independent.

\section{The charged Higgs boson}\label{Hp}


\begin{figure}[htbp]
  \begin{center}
    \epsfig{file=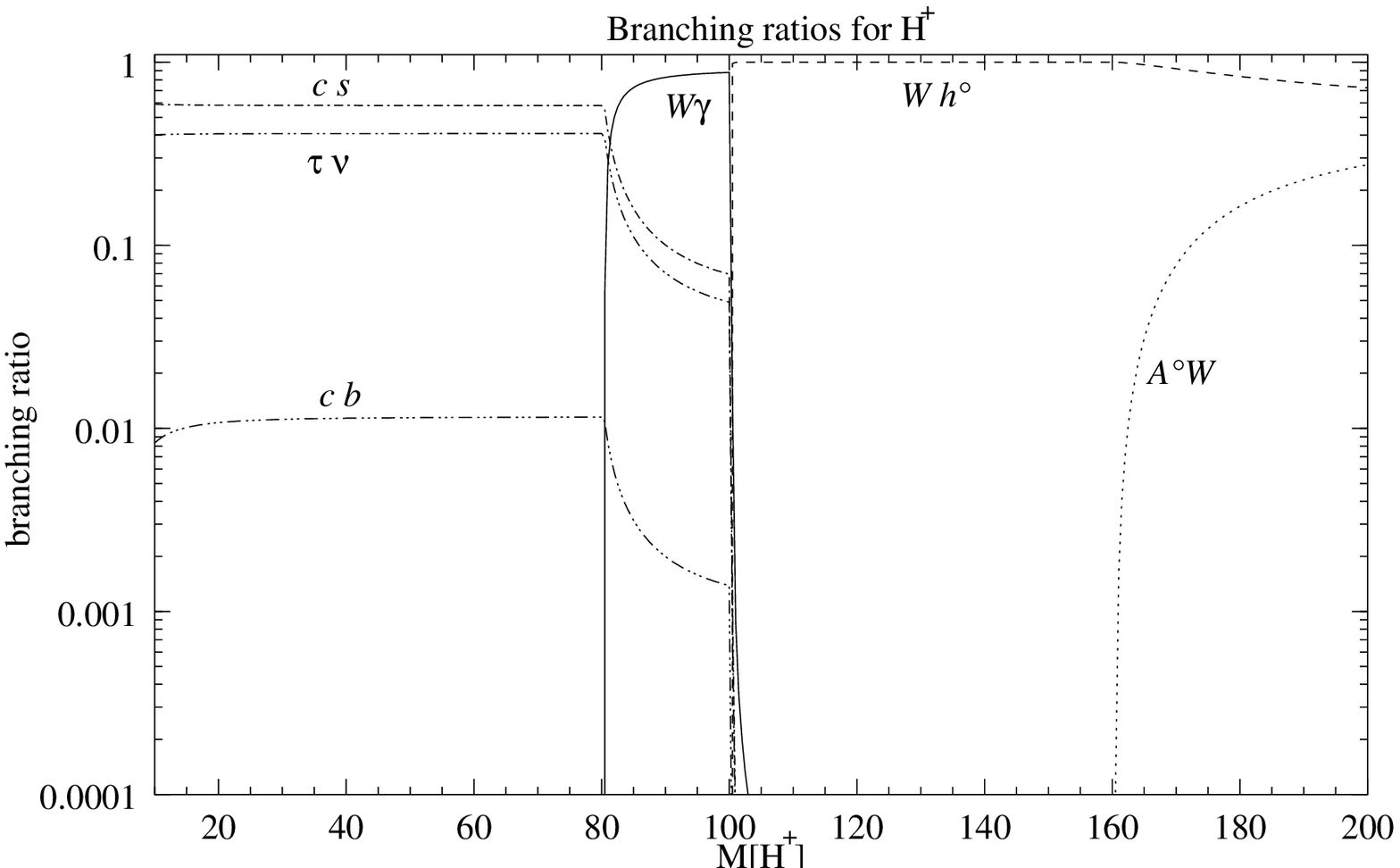,width=15cm}
    \caption{Branching ratios for $H^+$ at $\delta=0.01$, $m_{A^0}=80\ GeV$, 
      $m_{h^0}=20\ GeV$ and $m_{H^0}=300\ GeV$ in potential $A$.}
    \label{fig:HpA1}
  \end{center}
\end{figure}

\begin{figure}[htbp]
  \begin{center}
    \epsfig{file=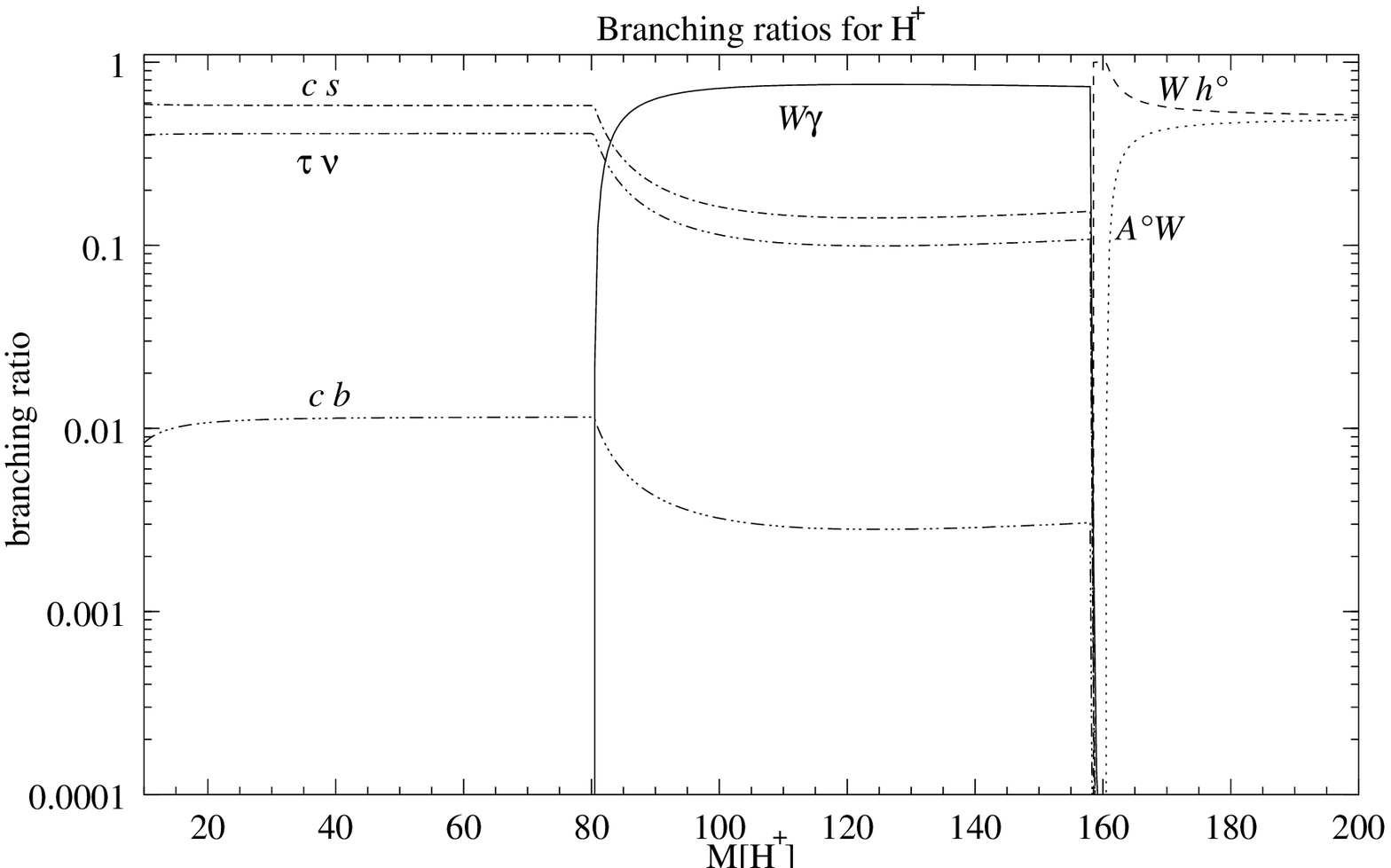,width=15cm}
    \caption{Branching ratios for $H^+$ at $\delta=0.01$, $m_{A^0}=80\ GeV$, 
      $m_{h^0}=78\ GeV$ and $m_{H^0}=500\ GeV$ in potential $B$.}
    \label{fig:HpB1}
  \end{center}
\end{figure}

\begin{figure}[htbp]
  \begin{center}
    \epsfig{file=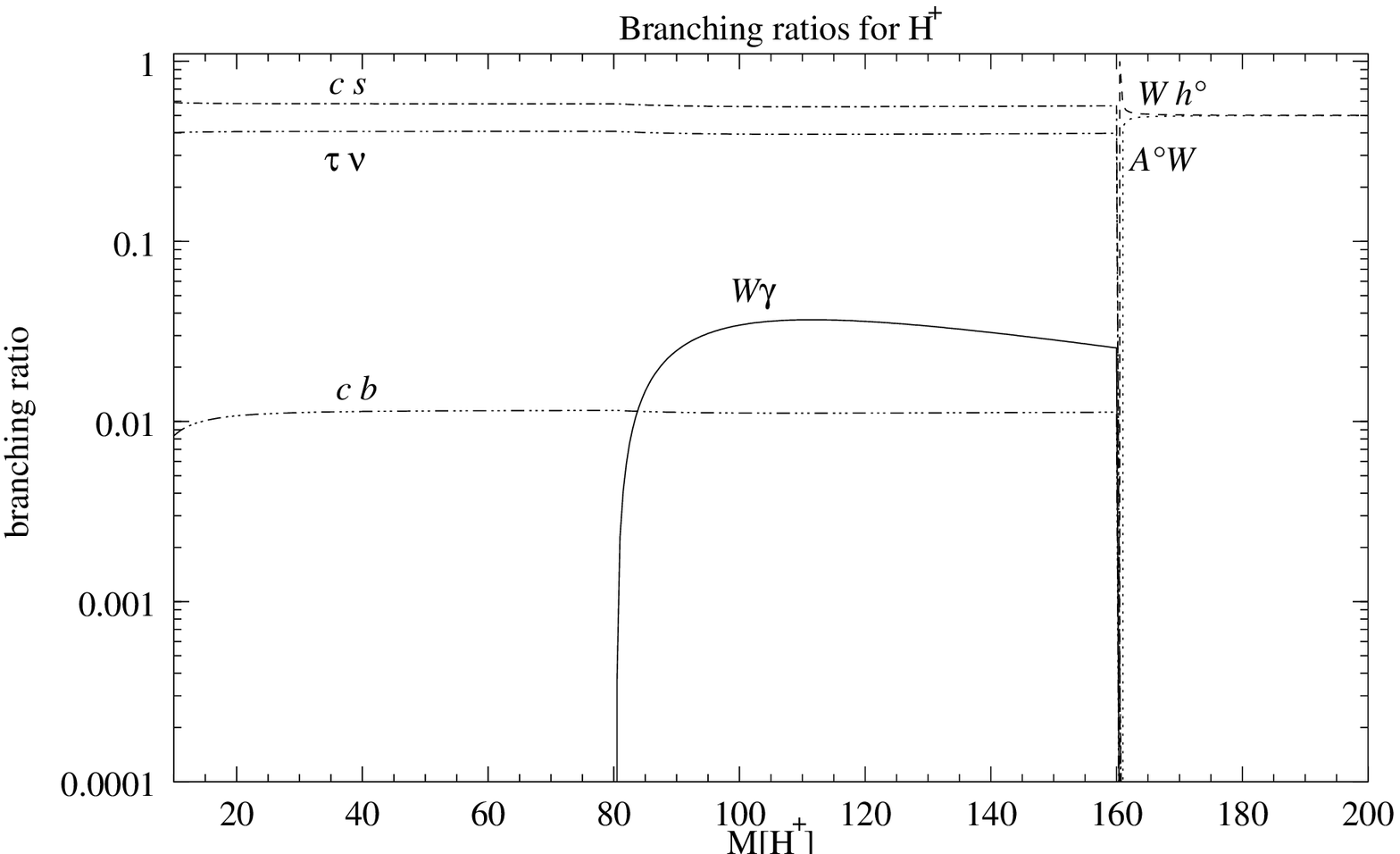,width=15cm}
    \caption{Branching ratios for $H^+$ at $\delta=0.01$, $m_{A^0}=80.1\ GeV$, 
      $m_{h^0}=80\ GeV$ and $m_{H^0}=500\ GeV$ in potential $B$.}
    \label{fig:HpB2}
  \end{center}
\end{figure}

The charged Higgs boson has the following tree-level decays:
\begin{displaymath}
  H^\pm \rightarrow f_{u,d} \bar{f}_{d,u} \quad ; \quad H^\pm \rightarrow W^\pm
  h^0 \quad ; \quad H^\pm \rightarrow W^\pm A^0 \quad ; \quad H^\pm
  \rightarrow W^\pm H^0 
\end{displaymath}
Moreover the following one-loop decays have to be considered:
\begin{displaymath}
  H^\pm \rightarrow W^\pm \gamma \quad ; \quad H^\pm \rightarrow W^\pm Z
\end{displaymath}
Again, the 16 (32) graphs for $H^\pm \rightarrow W^\pm \gamma$ ($H^\pm
\rightarrow W^\pm Z$) have been calculated with \xl \cite{xl1}.

In potential $A$ the branching ratios of the charged Higgs boson show
no surprises (fig.~\ref{fig:HpA1}). If the mass of $H^\pm$ is below
the $W$ mass the signature will be fermionic and independent of the
value of $\delta$. In this mass region the situation is similar to the
former situation concerning the pseudo-scalar Higgs boson. Decreasing
$\delta$ just decreases the total decay width, but leaves the
branching ratio unchanged. The situation changes, as soon as the
$W\gamma$ threshold is passed. Then, in the tiny $\delta$ region the
signature will be $H^\pm\rightarrow W^\pm\gamma$.  In the small
$\delta$ region, as $\delta$ grows the branching ratio of
$H^\pm\rightarrow W^\pm\gamma$ decreases to less than 1\%.
Consequently in the large $\delta$ region the signature is again
fermionic. As soon as decays to the $W$ and either $A^0$ or $h^0$ are
kinematically allowed, the sum of these decays will have an
approximately 100\% branching ratio for almost all values of $\delta$.
Only if $\delta$ becomes very large and $m_{H^+}>m_t+m_b$ the decay
$H^+\rightarrow t\bar{b}$ will be the dominant decay mode.

In potential $B$ below the $W\gamma$ and above the $Wh^0$ or $WA^0$
threshold the situation is the same as in potential $A$.
When the decay $H^\pm \rightarrow W^\pm \gamma$ is important, the
situation strongly depends on the choice of parameters. In principle,
due to the lack of an upper bound for $m_{h^0}$ in potential $B$, the
interval of $m_{H^+}$, where $H^\pm\rightarrow W^\pm\gamma$ could be
much larger than in potential $A$. On the other hand, it turns out
that due to the degeneracy of $m_{h^0}$ and $m_{A^0}$ the decay to
$W\gamma$ can be suppressed to a few percent in comparison to the
fermion decays (see fig.~\ref{fig:HpB2}). This behaviour can also be
seen for very tiny values of $\delta$. If the restriction on $m_{A^0}$
and $m_{h^0}$ is limbered and their masses just differ by a few $GeV$,
$H^\pm \rightarrow W^\pm\gamma$ regains its importance
(fig.~\ref{fig:HpB1}).  Moreover, in contrast to potential $A$, it can
still be the major decay in the small and at the start of the large
$\delta$ region. Even for $\delta=0.2$ the branching of
$H^\pm\rightarrow W^\pm\gamma$ can reach up to 10\%, if the masses of
the Higgs sector are chosen appropriately. Of course for an even
larger value of $\delta$ the branching ratio for this decay mode will
become unimportant.

\section{The heavy scalar Higgs boson}


\begin{figure}[htbp]
  \begin{center}
    \epsfig{file=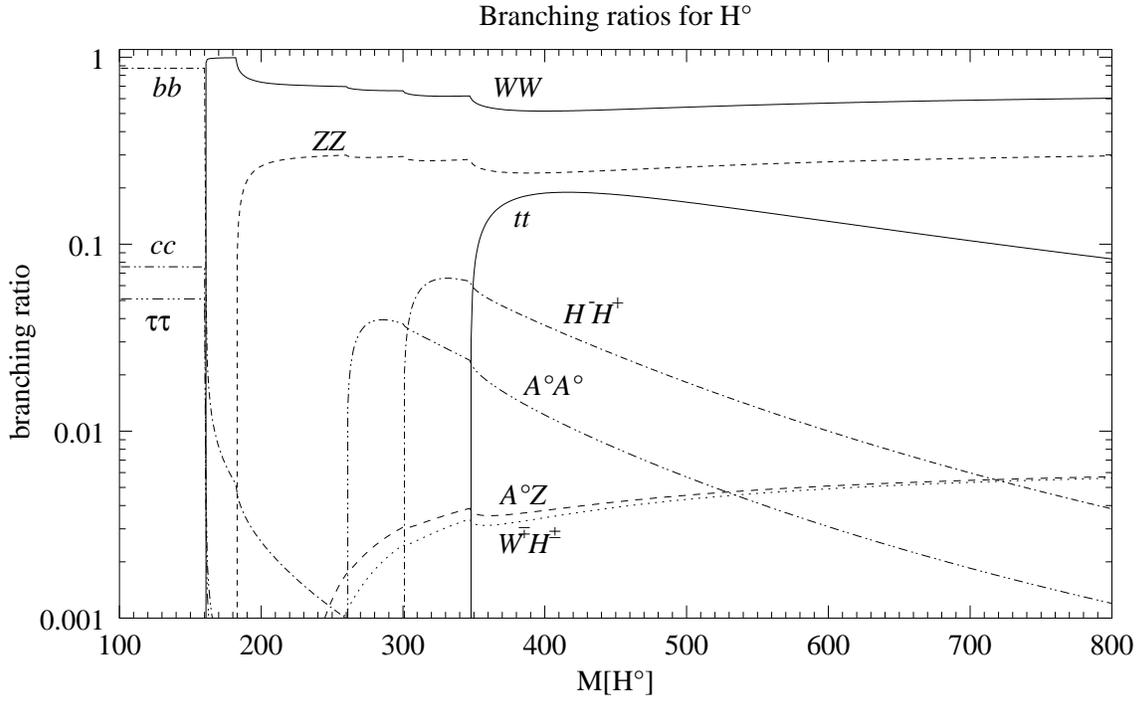,width=15cm}
    \caption{Branching ratios for $H^0$ at $\delta=0.1$ in potential $A$.}
    \label{fig:BigH}
  \end{center}
\end{figure}


\begin{figure}[htbp]
  \begin{center}
    \epsfig{file=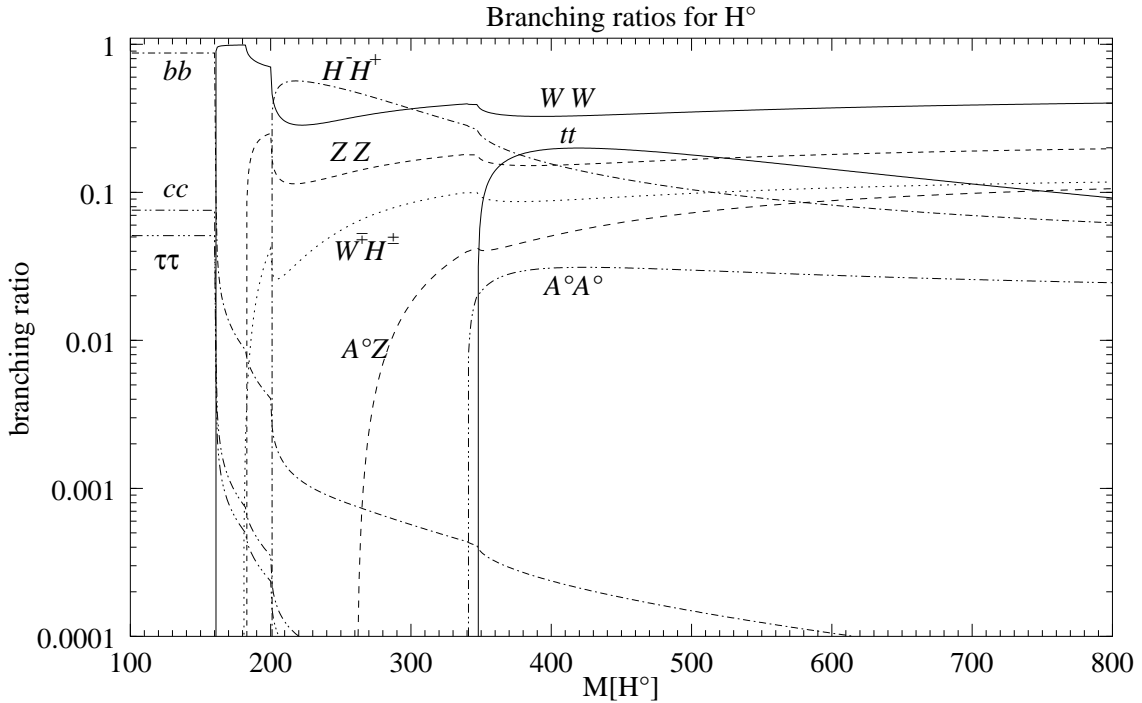,width=15cm}
    \caption{Branching ratios for $H^0$ at $\delta=0.5$ in potential $B$.}
    \label{fig:BigH2}
  \end{center}
\end{figure}

For the sake of completeness we show the decay modes of the heavy
scalar Higgs boson, although its coupling to the fermions is large.
Thus one-loop decays have no importance in the branching ratio of
$H^0$. So we just consider the following tree-level decays:
\begin{displaymath}
 \begin{array}{lclclclc}
   H^0 \rightarrow f \bar{f} & ; & 
    H^0 \rightarrow H^+ H^- & ; & 
    H^0 \rightarrow A^0 A^0 & ; &
    &  \\
    H^0 \rightarrow W^\pm H^\mp & ; &
    H^0 \rightarrow Z A^0 & ; &
    H^0 \rightarrow Z Z & ; &
    H^0 \rightarrow W^+ W^- & 
  \end{array}
\end{displaymath}
Note that the decay $H^0 \rightarrow h^0 h^0$ vanishes in the
fermiophobic limit (i.e. for $\alpha=\pi/2$).

A typical plot of the branching ratio as a function of the mass is shown
in fig.~\ref{fig:BigH}. Obviously, the heavy scalar Higgs boson mainly
decays into $b\bar{b}$ below, and into $WW$ above the two vector boson
threshold. This behaviour is typical for both potentials. The only
difference between the potentials can be recognized in the purely
scalar decay modes. In potential $A$ their contribution varies from
$0\%$ to $\approx 20\%$ depending on the parameters chosen. In
potential $B$ the decay $H^0\rightarrow H^+ H^-$ can be the major
decay mode for some values of $\delta$, $m_{H^+}$ and $m_A$, as can be
seen in fig.~\ref{fig:BigH2}.

Only if $\delta > 1.0$, $H^0$ mainly decays to
$t\bar{t}$.  Notice that the branching ratios shown in
fig.~\ref{fig:BigH} are similar those obtained for the SM Higgs boson,
if the scalar decays are ignored.

\section{Constraints on the models}\label{exlim}

\begin{figure}[htbp]
  \begin{center}
    \epsfig{file=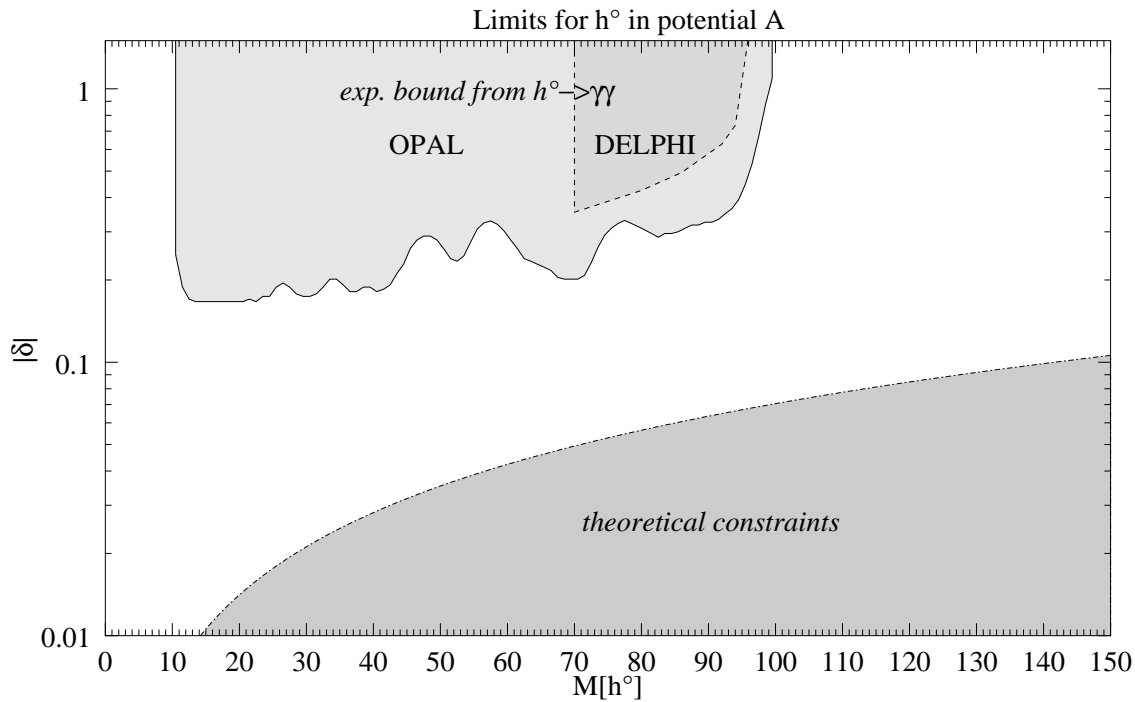,width=15cm}
    \caption{Bounds in the $m_{h^0}$-$\delta$ plane for
      potential $A$.}
    \label{fig:LimhA}
  \end{center}
\end{figure}

\begin{figure}[htbp]
  \begin{center}
    \epsfig{file=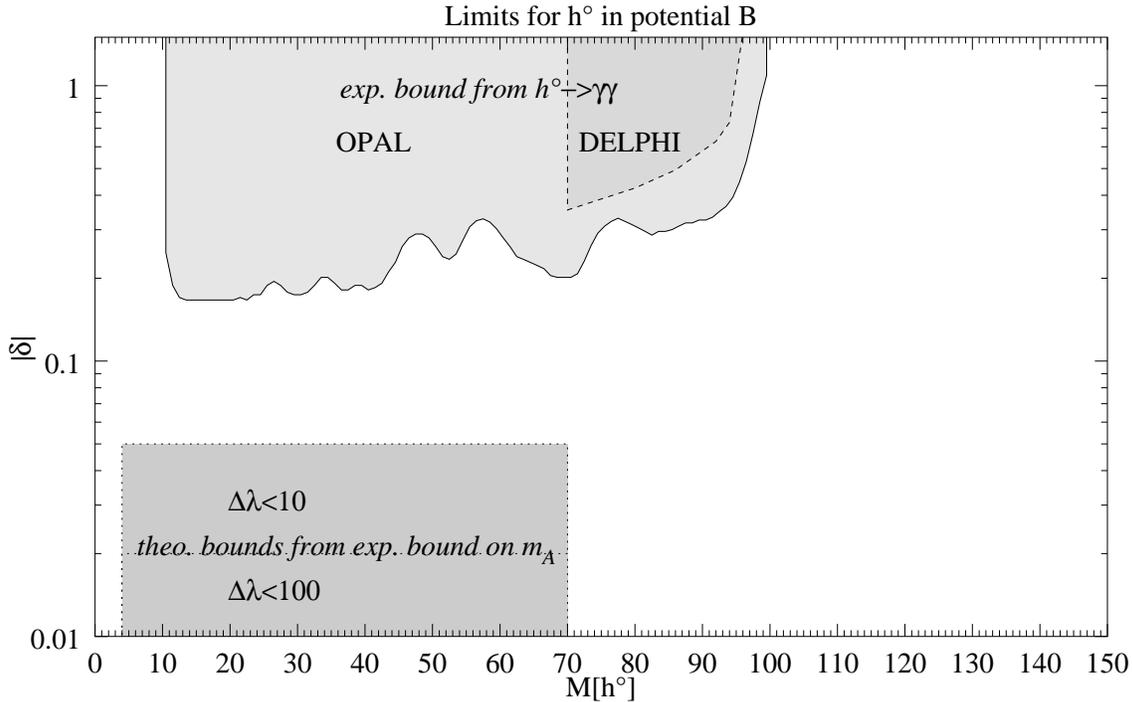,width=15cm}
    \caption{Bounds in the $m_{h^0}$-$\delta$ plane for
      potential $B$.}
    \label{fig:LimhB}
  \end{center}
\end{figure}

In this section we use the available experimental data and the bounds
derived in section \ref{masslim} to constrain the models.

Most production modes of the pseudo-scalar Higgs boson at LEP are
suppressed in the fermiophobic limit. An exception is the associated
production $Z^*\rightarrow h^0 A^0$ when kinematically allowed. The more
$\delta$ tends to zero the larger becomes the cross section for this
production mode. However, the obtained limit for $m_A$ is not independent of
the mass of the lightest scalar Higgs boson. This production mechanism
has recently been measured by the DELPHI coll. \cite{hex2}, where more
detailed results
can be found. For this associated production we roughly summarize their result in the following
inequation:
\begin{equation}
  \sqrt{m_{h^0}^2 + m_A^2} \ge 80\ GeV
\end{equation}

For the lightest scalar Higgs boson mass the most stringent bounds can
be derived from the experimental measurement of massive di-photon
resonances. The most recent data have been published in refs.
\cite{hex1,hex2}. We have used this data to exclude some regions in
the $m_{h^0}$-$\delta$ plane. We have plotted the results in
fig.~\ref{fig:LimhA} for potential $A$ and in fig.~\ref{fig:LimhB} for
potential $B$. 
Moreover we have inserted the
theoretical constraints shown in fig.~\ref{fig:Mhlimit1A}. In
fig.~\ref{fig:LimhA} (potential $A$) this can be seen as the lower
limit on $\delta$ for a given $h^0$ mass. For potential $B$ the
experimental bound on $m_A$ can be used to derive a lower limit on
$\delta$ for a given $m_{h^0}$. In fig.~\ref{fig:LimhB} we have
plotted this area for different values of
$\Delta\lambda$.\footnote{c.f.  section \ref{masslim}.}

Model independent bounds for the charged Higgs boson mass can be 
obtained using the universality of the electromagnetic coupling.
Measurements of $e^+ e^- \rightarrow H^+ H^-$ at LEP currently yield a
lower bound of $m_{H^+}>59\ GeV$ \cite{hex4}. In hadron colliders the
search for $t \rightarrow H^+ b$ gives a limit of $Br(t \rightarrow
H^+ b)\times Br(H^+\rightarrow\tau \nu_\tau) \le 50\%$ if $60\ GeV \le m_{H^+}\le
165 \ GeV$ \cite{hex3}. This imposes a lower as well as an upper limit
on $\beta$ in 2HDM's with coupling to fermions of type II or III.
Unfortunately, in the 2HDM I this simply gives a very large upper limit on
$\delta$, but no lower limit. Furthermore, we have shown in section
\ref{Hp} that for $m_{H^+}>m_W$ the decay $H^+ \rightarrow \tau
\nu_\tau$ can be suppressed in the fermiophobic limit.

\section{Conclusion and outlook}

We have calculated the branching ratios for all Higgs particles of
fermiophobic 2HDM´s as a function of the Higgs masses and $\delta$.
We have shown that the two different scalar sectors, models $A$ and
$B$, give rise to different signatures for some regions of the parameter
space. Most of the mass bounds based on a general 2HDM or on the MSSM
do not apply in the fermiophobic case. We have used the available experimental data
and tree-level unitarity bounds to constrain the models.  It turns
out, that there is still a wide region of this parameter space not yet
excluded by experimental data and still accessible at the LEP
collider. So, one should keep an open mind for surprises in the Higgs
sector.

\section{Acknowledgement}

We would like to thank our experimental colleagues at LIP for
inspiring discussions and A. Barroso for a careful reading of the
manuscript. L.B. is supported by JNICT under contract No.
BPD.16372.

\begin{appendix}

\section{Formulas for the decay widths}\label{Adw}

Here we present the most important formulas for the decay widths of
the Higgs particles. We use the Kallen function
$\lambda(x,y,z)=x^2+y^2+z^2-2xy-2xz-2yz$ in the formulas below. For
the lightest scalar the following tree-level decays have been
calculated:
\begin{eqnarray*}
 \Gamma (h^0\rightarrow W^+ W^-) & = & \frac{g^2\, m_W^2\sin^2\delta} 
     {8 \pi\, m_{h^0}^2}\,\sqrt{m_{h^0}^2-4 m_W^2}\,\left[\,
  1\,+\, \frac{\left(m_{h^0}^2-2m_W^2\right)^2}{8\,m_W^4} \,\right]\\
 \Gamma (h^0\rightarrow Z Z) & = & \frac{g^2\, m_Z^4\cos^2\delta} 
     {16 \pi\, m_W^2 m_{h^0}^2}\,\sqrt{m_{h^0}^2-4 m_Z^2}\,\left[\,
  1\,+\, \frac{\left(m_{h^0}^2-2m_Z^2\right)^2}{8\,m_Z^4} \,\right]\\
 \Gamma(h^0 \rightarrow Z A^0) & = & \frac{g^2 \cos^2\delta}{64 \pi\, m_W^2
    m_{h^0}^3}\lambda^{\frac{3}{2}}\left(m_{h^0}^2,m_{A^0}^2,m_Z^2\right) \\
 \Gamma (h^0 \rightarrow W^\pm H^\mp) & = & \frac{g^2\cos^2\delta}{64\pi\,
 m_{h^0}^3 m_W^2} \lambda^{\frac{3}{2}}(m_{h^0}^2,m_{H^+}^2,m_W^2) \\
  \Gamma^{(A/B)} (h^0\rightarrow A^0 A^0) & = &
    \frac{\sqrt{m_{h^0}^2-4m_{A^0}^2}}{32\pi\,m_{h^0}^2}\,
    \left|C^{(A/B)}_{[h^0 A^0 A^0]}\right|^2 \\
  \Gamma^{(A/B)} (h^0\rightarrow H^+ H^-) & = &
  \frac{\sqrt{m_{h^0}^2-4m_{H^+}^2}}{16\pi\,m_{h^0}^2}\, 
    \left|C^{(A/B)}_{[h^0 H^+ H^-]}\right|^2 \qquad\qquad . 
\end{eqnarray*}
The couplings $C^{(A/B)}_{[h^0 H^+ H^-]}$ for either potential $A$ or
potential $B$ are listed in appendix B. The one-loop decays have been
automatically calculated with \xl . Unfortunately the formulas are to
large to be shown here. A compact formula for $h^0 \rightarrow
\gamma\gamma$ in the MSSM can be found in ref.~\cite{Zerw1}.
  
For the pseudo-scalar Higgs boson one gets:
\begin{eqnarray*}
  \Gamma(A^0 \rightarrow f \bar{f}) & = & N_c \frac{g^2 m_f^2
  \cot^2\beta}{32 \pi m_W^2} \sqrt{\frac{m_{A^0}^2}{4}-m_f^2} \left[ 1
  - \frac{4 m_f^2}{m_{A^0}^2} \right] \\  
  \Gamma(A^0 \rightarrow Z h^0) & = & \frac{g^2 \cos^2\delta}{64 \pi\, m_W^2
  m_{A^0}^3}\lambda^{\frac{3}{2}}\left(m_{A^0}^2,m_{h^0}^2,m_Z^2\right) \\
 \Gamma(A^0 \rightarrow W^\pm H^\mp) & = & \frac{g^2}{64 \pi\, m_W^2
 m_{A^0}^3} \lambda^{\frac{3}{2}}\left(m_{A^0}^2,m_{H^+}^2,m_W^2\right) \\
  \Gamma(A^0 \rightarrow \gamma\gamma) & = & \frac{m_{A^0}^3}{32 \pi} \left|
  \frac{N_c{e}^{3}m_t^{2}\cot\beta}{9\,\sin\theta_W\,m_W \,
  {\pi }^{4}}
  \mbox{Oneloop3Pt}\left(0,0,0,m_{A^0},\halb m_{A^0},\halb m_{A^0},m_t,m_t,m_t\right) \right|^2
\end{eqnarray*}
$N_c$ denotes the number of quark colors. Beside $A^0 \rightarrow
\gamma\gamma$ all other formulas for one-loop decays have been
skipped. The definition of the OneLoop3Pt function
can be found in ref.~\cite{oneloop}. An ${\cal O}({\alpha}_{s})$
improved formula for $\Gamma(A^0 \rightarrow q \bar{q})$ can be found
in ref.~\cite{Zerw2}.

The charged Higgs boson has the following partial decay widths:
\begin{eqnarray*}
  \Gamma (H^+ \rightarrow f_t \bar{f_b}) & = & \frac{g^2\cot^2\beta \,N_c \left|
  V_{tb}\right|^2}{64\pi\,m_W^2}
  \frac{\lambda^{\frac{1}{2}}(m_{H^+}^2,m_{t}^2,m_b^2)}{m_{H^+}^3} \left[
   \left(m_{H^+}^2-m_{t}^2-m_b^2\right)\left(m_{t}^2+m_b^2\right)
     + 4\,m_t^2 m_b^2 \right] \\
 \Gamma (H^+ \rightarrow W^+ h^0) & = & \frac{g^2\cos^2\delta}{64\pi\,
 m_{H^+}^3 m_W^2} \lambda^{\frac{3}{2}}(m_{H^+}^2,m_{h^0}^2,m_W^2) \\
 \Gamma (H^+ \rightarrow W^+ A^0) & = & \frac{g^2}{64\pi\,
 m_{H^+}^3 m_W^2} \lambda^{\frac{3}{2}}(m_{H^+}^2,m_{A^0}^2,m_W^2) \\  
 \Gamma (H^+ \rightarrow W^+ H^0) & = & \frac{g^2\sin^2\delta}{64\pi\,
 m_{H^+}^3 m_W^2} \lambda^{\frac{3}{2}}(m_{H^+}^2,m_{H^0}^2,m_W^2)  
\end{eqnarray*}
Note that $\Gamma (H^+ \rightarrow f_t \bar{f_b})$ is also valid for
leptons with $m_{\nu} \equiv m_t=0$ and $N_c=1$.  Again
we skip the formula for $H^+ \rightarrow W^+ \gamma$ and $H^+
\rightarrow W^+ Z$ due to its length.

For the heavy Higgs boson ($H^0$) we calculate:
\begin{eqnarray*}
  \Gamma (H^0\rightarrow f\bar{f}) & = & \frac{g^2\,N_c\,m_f^2
  \,\sin^2\alpha}{64 \pi\, m_W^2 \sin^2\beta}\,\sqrt{m_{H^0}^2-4 m_f^2} \\
  \Gamma (H^0\rightarrow W^+ W^-) & = & \frac{g^2\, m_W^2\cos^2\delta} 
     {8 \pi\, m_{H^0}^2}\,\sqrt{m_{H^0}^2-4 m_W^2}\,\left[\,
  1\,+\, \frac{\left(m_{H^0}^2-2m_W^2\right)^2}{8\,m_W^4} \,\right]\\
  \Gamma (H^0\rightarrow Z Z) & = & \frac{g^2\, m_Z^4\cos^2\delta} 
     {16 \pi\, m_W^2 m_{H^0}^2}\,\sqrt{m_{H^0}^2-4 m_Z^2}\,\left[\,
  1\,+\, \frac{\left(m_{H^0}^2-2m_Z^2\right)^2}{8\,m_Z^4} \,\right]\\
  \Gamma(H^0 \rightarrow Z A^0) & = & \frac{g^2 \sin^2\delta}{64 \pi\, m_W^2
  m_{H^0}^3}\lambda^{\frac{3}{2}}\left(m_{H^0}^2,m_{A^0}^2,m_Z^2\right) \\
 \Gamma (H^0 \rightarrow W^\pm H^\mp) & = & \frac{g^2\sin^2\delta}{64\pi\,
    m_{H^0}^3 m_W^2} \lambda^{\frac{3}{2}}(m_{H^0}^2,m_{H^+}^2,m_W^2) \\
  \Gamma^{(A/B)} (H^0\rightarrow A^0 A^0) & = &
    \frac{\sqrt{m_{H^0}^2-4m_{A^0}^2}}{32\pi\,m_{H^0}^2}\,
    \left|C^{(A/B)}_{[H^0 A^0 A^0]}\right|^2 \\
  \Gamma^{(A/B)} (H^0\rightarrow H^+ H^-) & = &
  \frac{\sqrt{m_{H^0}^2-4m_{H^+}^2}}{16\pi\,m_{H^0}^2}\, 
    \left|C^{(A/B)}_{[H^0 H^+ H^-]}\right|^2 
\end{eqnarray*}

\section{Feynman rules}

In this section we present the Feynman rules for the triple and quartic
interactions of scalar fields which are different in both potentials.
A full treatment of all Feynman rules will be found in
ref.~\cite{rulesInPrep}.

We define the following quantities:\\
\parbox{8cm} {\begin{eqnarray*}
    A_{\alpha \beta} & \equiv & \cos^3 \beta \sin \alpha + \sin^3 \beta \cos \alpha \\
    B_{\alpha \beta} & \equiv & \cos^3 \beta \cos \alpha - \sin^3 \beta \sin \alpha \\
    C_{\alpha \beta} & \equiv & \sin^3 \alpha \cos \beta + \cos^3 \alpha \sin \beta \\
    D_{\alpha \beta} & \equiv & \cos^3 \alpha \cos \beta - \sin^3 \alpha \sin \beta \\
\end{eqnarray*}
}\hfill \parbox{8cm} {\begin{eqnarray*}
    E_{\alpha \beta} & \equiv & (\cos^2\alpha-\sin^2\beta)  \\
    F_{\alpha \beta} & \equiv & (\cos^2\alpha-\cos^2\beta)  \\
    G_{\alpha \beta} & \equiv & \cos \beta \sin \beta - 3 \cos \alpha \sin \alpha \\
    H_{\alpha \beta} & \equiv & \cos 2\beta \cos\delta\cos(\alpha+\beta) \\
    K_{\alpha \beta} & \equiv & \cos 2\beta (\cos^2 \alpha - \cos^2
    \beta)
\end{eqnarray*}
}

\subsection{Different triple scalar vertices in $V_A$}

\parbox{8cm} {\begin{tabular}{ll}
    $H^+ H^- h$ & $-\frac{ig}{M_W} \left( \frac{M_h^2}{\sin 2 \beta} B_{\alpha \beta}-M_{H^+}^2 \sin \delta \right)$\\ \\
    $H^+ H^- H$ & $-\frac{ig}{M_W} \left( \frac{M_H^2}{\sin 2 \beta} A_{\alpha \beta}+M_{H^+}^2 \cos \delta \right)$\\ \\
  $h H H$ &  $-\frac{ig}{2M_W} \frac{\sin 2 \alpha \sin \delta }{\sin 2 \beta}(2 M_H^2 + M_h^2)$\\ \\
  $h h H$ &  $-\frac{ig}{2M_W} \frac{\sin 2 \alpha \cos \delta }{\sin 2 \beta}(M_H^2 + 2 M_h^2)$  \\ \\
\end{tabular}
}\hfill \parbox{8cm} {\begin{tabular}{ll}
  $h h h$ &  $-\frac{3ig}{M_W} \frac{M_h^2}{\sin 2 \beta} D_{\alpha \beta}$\\ \\
  $H H H$ &  $-\frac{3ig}{M_W} \frac{M_H^2}{\sin 2 \beta} C_{\alpha \beta}$\\ \\
    $A A h$ & $-\frac{ig}{M_W} \left( \frac{M_h^2}{\sin 2 \beta} B_{\alpha \beta}-M_{A}^2 \sin \delta \right)$\\ \\
    $A A H$ & $-\frac{ig}{M_W} \left( \frac{M_H^2}{\sin 2 \beta} A_{\alpha \beta}+M_{A}^2 \cos \delta \right)$\\ \\
\end{tabular}
}

\subsection{Different triple scalar vertices in $V_B$}

\parbox{9cm} {\begin{tabular}{ll} $H^+ H^- h$ & $-\frac{ig}{M_W}
    \left( \frac{M_h^2}{\sin 2 \beta} B_{\alpha \beta}-M_{H^+}^2 \sin
      \delta -
      \frac{\cos\delta}{\sin 2\beta} M_A^2\right)$\\ \\
    $H^+ H^- H$ & $-\frac{ig}{M_W} \left( \frac{M_H^2}{\sin 2 \beta}
      A_{\alpha \beta}+M_{H^+}^2 \cos \delta +
      \frac{\sin\delta}{\sin 2\beta} M_A^2\right)$\\ \\
  $h H H$ & $-\frac{ig\sin \delta}{2M_W\sin 2 \beta} \left(\sin 2
    \alpha  (2 M_H^2 + M_h^2) + M_A^2 G_{\alpha\beta}\right)$\\ \\
  $h h H$ & $-\frac{ig\cos\delta}{2M_W\sin 2 \beta} \left(\sin 2
    \alpha (M_H^2 + 2 M_h^2)
    +M_A^2 G_{\alpha\beta} \right)$  \\ \\
\end{tabular}
}\hfill \parbox{7.5cm} {\begin{tabular}{ll} $h h h$ &
  $-\frac{3ig}{M_W\sin 2 \beta} \left(M_h^2 D_{\alpha \beta}
    - M_A^2 \,E_{\alpha \beta}\cos\delta \,\right)$\\ \\
  $H H H$ & $-\frac{3ig}{M_W\sin 2 \beta} \left( M_H^2 C_{\alpha
      \beta}
    +M_A^2 \, F_{\alpha \beta}\sin\delta \,\right)$\\ \\
    $A A h$ & $-\frac{ig}{M_W} \left( \frac{M_h^2}{\sin 2 \beta}
      B_{\alpha
        \beta}-M_{A}^2 \frac{\cos \delta \cos 2 \beta}{\sin 2 \beta} \right)$\\ \\
    $A A H$ & $-\frac{ig}{M_W} \left( \frac{M_H^2}{\sin 2 \beta}
      A_{\alpha
        \beta}-M_{A}^2 \frac{\sin \delta \cos 2 \beta}{\sin 2 \beta} \right)$\\ \\
\end{tabular}
}

\subsection{Different quartic scalar vertices for $V_A$}

\begin{tabular}{lcl}
$H^+ H^- H^+ H^-$ & & $-\frac{2\,ig^2}{\sin^2 2 \beta M_W^2}(M_H^2 A_{\alpha \beta}^2 + M_h^2 B_{\alpha \beta}^2)$ \\ \\
$A A A A$ & & $-\frac{3ig^2}{\sin^2 2 \beta M_W^2}(M_H^2 A_{\alpha \beta}^2 + M_h^2 B_{\alpha \beta}^2)$ \\ \\
$A A H^+ H^-$ & & $-\frac{ig}{\sin^2 2 \beta M_W^2}(M_H^2 A_{\alpha \beta}^2 + M_h^2 B_{\alpha \beta}^2)$ \\ \\
$H^+ H^- h h$ & & $-\frac{ig^2}{2 M_W^2}\left[ \frac{1}{\sin^2 2 \beta}(M_H^2 A_{\alpha \beta} \sin 2 \alpha \cos \delta + 2 M_h^2 B_{\alpha \beta}D_{\alpha \beta}) + M_{H^+}^2 \sin^2 \delta \right]$ \\ \\
$H^+ H^- H H$ & & $-\frac{ig^2}{2 M_W^2}\left[ \frac{1}{\sin^2 2 \beta}(2 M_H^2 A_{\alpha \beta} C_{\alpha \beta} + M_h^2 B_{\alpha \beta}\sin 2 \alpha \sin \delta) + M_{H^+}^2 \cos^2 \delta \right]$ \\ \\
$A A h h$ & & $-\frac{ig^2}{2 M_W^2}\left[ \frac{1}{\sin^2 2 \beta}(M_H^2 A_{\alpha \beta} \sin 2 \alpha \cos \delta + 2 M_h^2 B_{\alpha \beta}D_{\alpha \beta}) + M_{A}^2 \sin^2 \delta \right]$ \\ \\
$A A H H$ & & $-\frac{ig^2}{2 M_W^2}\left[ \frac{1}{\sin^2 2 \beta}(2 M_H^2 A_{\alpha \beta} C_{\alpha \beta} + M_h^2 B_{\alpha \beta}\sin 2 \alpha \sin \delta) + M_{A}^2 \cos^2 \delta \right]$ \\ \\
$H^+ H^- H h$ & & $-\frac{ig^2}{2 M_W^2}\left[ \frac{\sin 2 \alpha}{\sin^2 2
    \beta}(M_H^2 A_{\alpha \beta} \sin \delta + M_h^2
  B_{\alpha \beta} \cos \delta) - \frac{1}{2} M_{H^+}^2 \sin 2 \delta \right]$ \\ \\
$A A H h$ & & $-\frac{ig^2}{2 M_W^2}\left[ \frac{\sin 2 \alpha}{\sin^2 2 \beta}(M_H^2 A_{\alpha \beta} \sin \delta + M_h^2 B_{\alpha \beta}\cos \delta) - \frac{1}{2} M_{A}^2 \sin 2 \delta \right]$ \\ \\
$h h h h$ & & $-\frac{3\,ig^2}{\sin^2 2 \beta M_w^2} (4 M_h^2 D_{\alpha \beta}^2 + M_H^2 \sin^2 2 \alpha \cos^2 \delta)$ \\ \\
$H H H H$ & & $-\frac{3\,ig^2}{\sin^2 2 \beta M_W^2} (M_h^2 \sin^2 2 \alpha \sin^2 \delta + 4 M_H^2 C_{\alpha \beta}^2)$ \\ \\
$h h h H$ & & $-\frac{3\,ig^2}{2 \sin^2 2 \beta M_W^2} (4 M_h^2 D_{\alpha \beta} \sin 2 \alpha \cos \delta + M_H^2 \sin^2 2 \alpha \sin 2 \delta)$ \\ \\
$H H H h$ & & $-\frac{3ig^2}{2 \sin^2 2 \beta M_W^2} (M_h^2 \sin^2 2 \alpha \sin 2 \delta + 4 M_H^2 C_{\alpha \beta} \sin 2 \alpha \sin \delta)$ \\ \\
$h h H H$ & & $- \frac{i g^2 \sin 2 \alpha}{4 \sin 2 \beta M_W^2} \left[ M_H^2 - M_h^2 + \frac{3 \sin 2 \alpha}{\sin 2 \beta} (\sin^2 \delta M_H^2 + \cos^2 \delta M_h^2) \right]$ \\ \\
$A A G^0 G^0$ & & $- \frac{ig^2}{4 M_W^2} \left[ \frac{\sin 2 \alpha}{\sin 2 \beta}( M_H^2 - M_h^2) + 3 (\sin^2 \delta M_H^2 + \cos^2 \delta M_h^2) \right]$ \\ \\
$H^+ H^- G^+ G^-$ & & $- \frac{ig^2}{4 M_W^2} \left[ M_A^2 + \frac{\sin 2 \alpha}{\sin 2 \beta}( M_H^2 - M_h^2) + 2 (\sin^2 \delta M_H^2 + \cos^2 \delta M_h^2) \right]$ \\ \\
\end{tabular}


\begin{tabular}{lcl}
$G^+ G^- A A$ & & $- \frac{ig^2}{2 M_W^2} \left[ M_{H^+}^2 + \frac{1}{\sin 2 \beta}(\cos \delta A_{\alpha \beta}M_H^2 - \sin \delta  B_{\alpha \beta} M_h^2) \right]$ \\ \\
$H^+ H^- G^0 G^0$ & & $- \frac{ig^2}{2 M_W^2} \left[ M_{H^+}^2 + \frac{1}{\sin 2 \beta}(\cos \delta A_{\alpha \beta}M_H^2 - \sin \delta  B_{\alpha \beta} M_h^2) \right]$ \\ \\
$H^+ H^- H^{\mp}G^{\pm}$ & & $- \frac{ig^2}{M_W^2\sin 2 \beta} \left[
  \sin \delta A_{\alpha \beta}M_H^2 + \cos \delta B_{\alpha \beta} M_h^2 \right]$ \\ \\
$H^+ H^- G^0 A$ & & $- \frac{ig^2}{2 M_W^2\sin 2 \beta} \left[
  \sin \delta A_{\alpha \beta}M_H^2 + \cos \delta B_{\alpha \beta} M_h^2 \right]$ \\ \\
$A A A G^0$ & & $- \frac{3\,ig^2}{2 M_W^2\sin 2 \beta} \left[
  \sin \delta A_{\alpha \beta}M_H^2 + \cos \delta B_{\alpha \beta} M_h^2 \right]$ \\ \\
$A A H^{\mp}G^{\pm}$ & & $- \frac{ig^2}{2 M_W^2\sin 2 \beta} \left[
  \sin \delta A_{\alpha \beta}M_H^2 + \cos \delta B_{\alpha \beta} M_h^2 \right]$ \\ \\
$G^+ G^- h h$ & & $- \frac{ig^2}{4 M_W^2} \left[ \frac{1}{\sin 2 \beta}(\sin 2 \alpha \cos^2 \delta M_H^2 - 2 \sin \delta D_{\alpha \beta} M_h^2) + 2 \cos^2 \delta M_{H^+}^2 \right]$ \\ \\
$G^0 G^0 h h$ & & $- \frac{ig^2}{4 M_W^2} \left[ \frac{1}{\sin 2 \beta}(\sin 2 \alpha \cos^2 \delta M_H^2 - 2 \sin \delta D_{\alpha \beta} M_h^2) + 2 \cos^2 \delta M_{A}^2 \right]$ \\ \\
$G^+ G^- H H$ & & $- \frac{ig^2}{4 M_W^2} \left[ \frac{1}{\sin 2 \beta}(2 \cos \delta C_{\alpha \beta} M_H^2 - \sin^2 \delta \sin 2 \alpha M_h^2) + 2 \sin^2 \delta M_{H^+}^2 \right]$ \\ \\
$G^0 G^0 H H$ & & $- \frac{ig^2}{4 M_W^2} \left[ \frac{1}{\sin 2 \beta}(2 \cos \delta C_{\alpha \beta} M_H^2 - \sin^2 \delta \sin 2 \alpha M_h^2) + 2 \sin^2 \delta M_{A}^2 \right]$ \\ \\
$H^{\mp}G^{\pm}H H$ & & $- \frac{ig^2}{8 M_W^2} \left[ \frac{1}{\sin 2 \beta}(4 \sin \delta C_{\alpha \beta} M_H^2 + \sin 2 \delta \sin 2 \alpha M_h^2) - 2 \sin 2 \delta M_{H^+}^2 \right]$ \\ \\
$H^{\mp}G^{\pm}h h$ & & $- \frac{ig^2}{8 M_W^2} \left[ \frac{1}{\sin 2 \beta}( \sin 2 \delta \sin 2 \alpha M_H^2 + 4 \cos \delta D_{\alpha \beta} M_h^2) + 2 \sin 2 \delta M_{H^+}^2 \right]$ \\ \\
$A G^0 H H$ & & $- \frac{ig^2}{8 M_W^2} \left[ \frac{1}{\sin 2 \beta}(4 \sin \delta C_{\alpha \beta} M_H^2 + \sin 2 \delta \sin 2 \alpha M_h^2) - 2 \sin 2 \delta M_{A}^2 \right]$ \\ \\
$A G^0 h h$ & & $- \frac{ig^2}{8 M_W^2} \left[ \frac{1}{\sin 2 \beta}(\sin 2 \alpha \sin 2 \delta M_H^2 + 4 \cos \delta D_{\alpha \beta} M_h^2) + 2 \sin 2 \delta M_{A}^2 \right]$ \\ \\
$G^+ G^- h H$ & & $- \frac{ig^2\sin 2\delta}{8 M_W^2} \left[ \frac{\sin 2 \alpha}{\sin 2 \beta}(M_H^2 - M_h^2) + 2 M_{H^+}^2 \right]$ \\ \\
$G^0 G^0 h H$ & & $- \frac{ig^2\sin 2\delta}{8 M_W^2} \left[ \frac{\sin 2 \alpha}{\sin 2 \beta}(M_H^2 - M_h^2) + 2 M_{A}^2 \right]$ \\ \\
%
%
$G^{\mp} H^{\pm} h H$ & & $- \frac{ig^2}{4 M_W^2} \left[ \frac{\sin 2 \alpha}{\sin 2 \beta}(\sin^2 \delta M_H^2 + \cos^2 \delta M_h^2) - \cos 2 \delta M_{H^+}^2 \right]$ \\ \\
$A G^0 h H$ & & $- \frac{ig^2}{4 M_W^2} \left[ \frac{\sin 2 \alpha}{\sin 2 \beta}(\sin^2 \delta M_H^2 + \cos^2 \delta M_h^2) - \cos 2 \delta M_{A}^2 \right]$ \\ \\
\end{tabular}

\subsection{Different quartic scalar vertices for $V_B$}

\begin{tabular}{lcl}
$H^+ H^- H^+ H^-$ & & $-\frac{2\,ig^2}{\sin^2 2 \beta M_W^2}(M_H^2
A_{\alpha \beta}^2 + M_h^2 B_{\alpha \beta}^2 - M_A^2\cos^2 2\beta)$ \\ \\
$A A A A$ & & $-\frac{3ig^2}{\sin^2 2 \beta M_W^2}(M_H^2 A_{\alpha \beta}^2 + M_h^2 B_{\alpha \beta}^2 - M_A^2\cos^2 2\beta)$ \\ \\
\end{tabular}

\newpage

\begin{tabular}{lcl}
$A A H^+ H^-$ & & $-\frac{ig}{\sin^2 2 \beta M_W^2}(M_H^2 A_{\alpha \beta}^2 + M_h^2 B_{\alpha \beta}^2 - M_A^2\cos^2 2\beta)$ \\ \\
$H^+ H^- h h$ & & $-\frac{ig^2}{2 M_W^2}\left[ \frac{1}{\sin^2 2
    \beta}(M_H^2 A_{\alpha \beta} \sin 2 \alpha \cos \delta + 2 M_h^2
  B_{\alpha \beta}D_{\alpha \beta} - H_{\alpha \beta}M_A^2 ) + \left(M_{H^+}^2-M_{A}^2\right) \sin^2 \delta \right]$ \\ \\
$H^+ H^- H H$ & & $-\frac{ig^2}{2 M_W^2}\left[ \frac{1}{\sin^2 2
    \beta}(2 M_H^2 A_{\alpha \beta} C_{\alpha \beta} + M_h^2 B_{\alpha
    \beta}\sin 2 \alpha \sin \delta - K_{\alpha\beta} M_A^2) + \left(M_{H^+}^2 -M_A^2\right)\cos^2 \delta \right]$ \\ \\
$A A h h$ & & $-\frac{ig^2}{2 M_W^2\sin^2 2 \beta}\left[M_H^2
  A_{\alpha \beta} \sin 2 \alpha \cos \delta + 2 M_h^2 B_{\alpha
    \beta}D_{\alpha \beta} - H_{\alpha\beta} M_{A}^2  \right]$ \\ \\
$A A H H$ & & $-\frac{ig^2}{2 M_W^2\sin^2 2 \beta}\left[2 M_H^2
  A_{\alpha \beta} C_{\alpha \beta} + M_h^2 B_{\alpha \beta}\sin 2
  \alpha \sin \delta - K_{\alpha\beta} M_{A}^2 \right]$ \\ \\
$H^+ H^- H h$ & & $-\frac{ig^2}{2 M_W^2}\left[ \frac{\sin 2 \alpha}{\sin^2 2
    \beta}(M_H^2 A_{\alpha \beta} \sin \delta + M_h^2
  B_{\alpha \beta}\cos \delta -M_A^2\cos 2 \beta) - \frac{1}{2} (M_{H^+}^2 -M_A^2)\sin 2 \delta \right]$ \\ \\
$A A H h$ & & $-\frac{ig^2}{2 M_W^2} \frac{\sin 2 \alpha}{\sin^2
    2 \beta}\left[M_H^2 A_{\alpha \beta} \sin \delta + M_h^2 B_{\alpha
    \beta}\cos \delta -  M_{A}^2 \cos 2 \beta \right]$ \\ \\
$h h h h$ & & $-\frac{3\,ig^2}{\sin^2 2 \beta M_w^2} \left(4 M_h^2
  D_{\alpha \beta}^2 + M_H^2 \sin^2 2 \alpha \cos^2 \delta - M_A^2
E_{\alpha\beta}^2\right)$ \\ \\
$H H H H$ & & $-\frac{3\,ig^2}{\sin^2 2 \beta M_W^2} \left(M_h^2 \sin^2 2
\alpha \sin^2 \delta + 4 M_H^2 C_{\alpha \beta}^2 - M_A^2
F_{\alpha\beta}^2\right)$ \\ \\
$h h h H$ & & $-\frac{3\,ig^2\sin 2 \alpha}{2 \sin^2 2 \beta M_W^2} \left[4
M_h^2 D_{\alpha \beta} \cos \delta + M_H^2 \sin 2\alpha \sin 2 \delta
-M_A^2 E_{\alpha\beta}\right]$ \\ \\
$H H H h$ & & $-\frac{3ig^2\sin 2\alpha}{2 \sin^2 2 \beta M_W^2} \left[M_h^2 \sin 2
\alpha \sin 2 \delta + 4 M_H^2 C_{\alpha \beta} \sin
\delta)-M_A^2(\sin^2\alpha-\sin^2\beta)\right]$ \\ \\
$h h H H$ & & $- \frac{i g^2 \sin 2 \alpha}{4 \sin 2 \beta M_W^2}
\left[ M_H^2 - M_h^2 + M_A^2\frac{3 \sin 2 \alpha}{\sin 2 \beta}
  (\sin^2 \delta M_H^2 + \cos^2 \delta M_h^2) -M_A^2 \right]$ \\ \\
$A A G^0 G^0$ & & $- \frac{ig^2}{4 M_W^2} \left[ \frac{\sin 2
    \alpha}{\sin 2 \beta}( M_H^2 - M_h^2) + 3 (\sin^2 \delta M_H^2 +
  \cos^2 \delta M_h^2)  - 2\,M_A^2\right]$ \\ \\
$H^+ H^- G^+ G^-$ & & $- \frac{ig^2}{4 M_W^2} \left[ 
  \frac{\sin 2 \alpha}{\sin 2 \beta}( M_H^2 - M_h^2) + 2 (\sin^2
  \delta M_H^2 + \cos^2 \delta M_h^2)\right]$ \\ \\
$G^+ G^- A A$ & & $- \frac{ig^2}{2 M_W^2} \left[ M_{H^+}^2 -M_A^2 + \frac{1}{\sin 2 \beta}(\cos \delta A_{\alpha \beta}M_H^2 - \sin \delta  B_{\alpha \beta} M_h^2) \right]$ \\ \\
$H^+ H^- G^0 G^0$ & & $- \frac{ig^2}{2 M_W^2} \left[ M_{H^+}^2 -M_A^2 + \frac{1}{\sin 2 \beta}(\cos \delta A_{\alpha \beta}M_H^2 - \sin \delta  B_{\alpha \beta} M_h^2) \right]$ \\ \\
$H^+ H^- H^{\mp}G^{\pm}$ & & $- \frac{ig^2}{M_W^2\sin 2 \beta} \left[
  \sin \delta A_{\alpha \beta}M_H^2 + \cos \delta B_{\alpha \beta}
  M_h^2-M_A^2\cos 2 \beta \right]$ \\ \\
$H^+ H^- G^0 A$ & & $- \frac{ig^2}{2\,M_W^2\sin 2 \beta} \left[
  \sin \delta A_{\alpha \beta}M_H^2 + \cos \delta B_{\alpha \beta} M_h^2-M_A^2\cos 2 \beta \right]$ \\ \\
$A A A G^0$ & & $- \frac{3\,ig^2}{2\,M_W^2\sin 2 \beta} \left[
  \sin \delta A_{\alpha \beta}M_H^2 + \cos \delta B_{\alpha \beta} M_h^2-M_A^2\cos 2 \beta \right]$ \\ \\
$A A H^{\mp}G^{\pm}$ & & $- \frac{ig^2}{2\,M_W^2\sin 2 \beta} \left[
  \sin \delta A_{\alpha \beta}M_H^2 + \cos \delta B_{\alpha \beta} M_h^2-M_A^2\cos 2 \beta\right]$ \\ \\
$G^+ G^- h h$ & & $- \frac{ig^2}{4 M_W^2} \left[ \frac{1}{\sin 2 \beta}(\sin 2 \alpha \cos^2 \delta M_H^2 - 2 \sin \delta D_{\alpha \beta} M_h^2) + 2 \cos^2 \delta (M_{H^+}^2 -M_A^2)\right]$ \\ \\
$G^0 G^0 h h$ & & $- \frac{ig^2}{4 M_W^2} \left[ \frac{1}{\sin 2 \beta}(\sin 2 \alpha \cos^2 \delta M_H^2 - 2 \sin \delta D_{\alpha \beta} M_h^2) \right]$ \\ \\
\end{tabular}


\begin{tabular}{lcl}
$G^+ G^- H H$ & & $- \frac{ig^2}{4 M_W^2} \left[ \frac{1}{\sin 2
  \beta}(2 \cos \delta C_{\alpha \beta} M_H^2 - \sin^2 \delta \sin 2
  \alpha M_h^2) + 2 \sin^2 \delta (M_{H^+}^2 -M_A^2) \right]$ \\ \\
$G^0 G^0 H H$ & & $- \frac{ig^2}{4 M_W^2} \left[ \frac{1}{\sin 2 \beta}(2 \cos \delta C_{\alpha \beta} M_H^2 - \sin^2 \delta \sin 2 \alpha M_h^2)  \right]$ \\ \\
$H^{\mp}G^{\pm}H H$ & & $- \frac{ig^2}{8 M_W^2} \left[ \frac{1}{\sin 2
  \beta}(4 \sin \delta C_{\alpha \beta} M_H^2 + \sin 2 \delta \sin 2
  \alpha M_h^2 + 2\,M_A^2 F_{\alpha\beta} ) - 2 \sin 2 \delta (M_{H^+}^2 -M_A^2) \right]$ \\ \\
$H^{\mp}G^{\pm}h h$ & & $- \frac{ig^2}{8 M_W^2} \left[ \frac{1}{\sin 2 \beta}( \sin 2 \delta \sin 2 \alpha M_H^2 + 4 \cos \delta D_{\alpha \beta} M_h^2+ 2\,M_A^2 E_{\alpha\beta}) + 2 \sin 2 \delta (M_{H^+}^2-M_A^2) \right]$ \\ \\
$A G^0 H H$ & & $- \frac{ig^2}{8 M_W^2} \left[ \frac{1}{\sin 2 \beta}(4 \sin \delta C_{\alpha \beta} M_H^2 + \sin 2 \delta \sin 2 \alpha M_h^2+ 2\,M_A^2 E_{\alpha\beta}) \right]$ \\ \\
$A G^0 h h$ & & $- \frac{ig^2}{8 M_W^2} \left[ \frac{1}{\sin 2 \beta}(\sin 2 \alpha \sin 2 \delta M_H^2 + 4 \cos \delta D_{\alpha \beta} M_h^2+ 2\,M_A^2 F_{\alpha\beta}) \right]$ \\ \\
$G^+ G^- h H$ & & $- \frac{ig^2\sin 2\delta}{8 M_W^2} \left[
  \frac{\sin 2 \alpha}{\sin 2 \beta}(M_H^2 - M_h^2) + 2 ( M_{H^+}^2
   - M_A^2) \right]$ \\ \\
$G^0 G^0 h H$ & & $- \frac{ig^2\sin 2\delta}{8 M_W^2} \left[ \frac{\sin 2 \alpha}{\sin 2 \beta}(M_H^2 - M_h^2) \right]$ \\ \\
%
%
$G^{\mp} H^{\pm} h H$ & & $- \frac{ig^2}{4 M_W^2} \left[ \frac{\sin 2 \alpha}{\sin 2 \beta}(\sin^2 \delta M_H^2 + \cos^2 \delta M_h^2-M_A^2) - \cos 2 \delta (M_{H^+}^2-M_A^2) \right]$ \\ \\
$A G^0 h H$ & & $- \frac{ig^2}{4 M_W^2} \left[ \frac{\sin 2 \alpha}{\sin 2 \beta}(\sin^2 \delta M_H^2 + \cos^2 \delta M_h^2) \right]$ \\ \\
\end{tabular}

\end{appendix}


\begin{thebibliography}{10}

\bibitem{SMH}
The OPAL coll.
\newblock {\em Search for Neutral Higgs bosons in $e^+e^-$ Collisions
  at $\sqrt{s}\approx 189$ GeV}.
\newblock {\em OPAL PN382} (1999).

\bibitem{Sant3}
{J. Velhinho, R. Santos and A. Barroso}.
\newblock {\em Phys. Lett.} {\bf B 322} (1994) 213--218.

\bibitem{bra1}
G.C. Branco and M.N. Rebelo.
\newblock {\em Phys. Lett.} {\bf B 160} (1985) 117.

\bibitem{Gun}
{J. F. Gunion, H. E. Haber, G. Kane and S. Dawson}.
\newblock {\em {The Higgs Hunter's Guide}}.\\
\newblock {Addison Wesley (Reading,MA,1990)}.

\bibitem{Sant4}
A. Barroso, L. Br\"ucher and R. Santos.
\newblock {\em Phys. Rev.} {\bf D 60} (1999) 035005.

\bibitem{Sant1}
{R. Santos and A. Barroso}.
\newblock {\em Phys. Rev.} {\bf D 56} (1997) 5366--5385.

\bibitem{SB1}
D. Kominis and R.S. Chivukula.
\newblock {\em Phys.Lett.} {\bf B 304} (1993) 152;
H. Komatsu.
\newblock {\em Prog.Theor.Phys.} {\bf 67} (1982) 1177;
R.A. Flores and M. Sher.
\newblock {\em Ann.Phys.(NY)} {\bf 148} (1983) 295;
S. Nie and M. Sher.
\newblock {\em Phys.Lett.} {\bf B 449} (1999) 89;
S. Kanemura, T. Kasai and Y. Okada.
\newblock {\em hep-ph} {\bf 9903289}.

\bibitem{trunal1}
{S. Kanemura, T. Kubota and E. Takasugi}.
\newblock {\em Phys. Lett.} {\bf B 313} (1993) 155--160.

\bibitem{trunal2}
{J. Maalampi, J. Sirkka and I. Vilja}.
\newblock {\em Phys. Lett.} {\bf B 265} (1991) 371--376.

\bibitem{xl1}
{L. Br\"ucher, J. Franzkowski and D. Kreimer}.
\newblock {\em Nucl.Instrum.Meth.} {\bf A 389} (1997) 323--342;
{L. Br\"ucher, J. Franzkowski and D. Kreimer}.
\newblock {\em hep-ph} {\bf 9710484};
{L. Br\"ucher, J. Franzkowski and D. Kreimer}.
\newblock {\em Comp. Phys. Comm.} {\bf 115} (1998) 140--160.

\bibitem{andr}
Jorge C. Rom\~ao and Sofia Andringa.
\newblock {\em Eur. Phys. J.} {\bf C7} (1999) 631.

\bibitem{Akr}
A.G.~Akeroyd,
{\em Nucl. Phys.} {\bf B 544} (1999) 557.

\bibitem{hex2}
{DELPHI coll.}.
\newblock {\em Search for non fermionic neutral Higgs couplings at LEP
  2}.
\newblock {\em Conference contribution to HEP Conference in
  Helsinki} (1999).

\bibitem{hex1}
{OPAL coll.}.
\newblock {\em Eur. Phys. J.} {\bf C 1} (1998) 31--43;
{ OPAL coll.}.
{\tt hep-ex/9907060}.

\bibitem{hex4}
{ALEPH coll.}.
\newblock {\tt hep-ex/9902031}.

\bibitem{hex3}
F. M. Borzumati and A. Djouadi.
\newblock {\tt hep-ph/9806301}.

\bibitem{Zerw1}
{M. Spira, A. Djouadi, D. Graudenz, P. M. Zerwas}.
\newblock {\em Nuc. Phys.} {\bf B 453} (1995) 17--82.

\bibitem{oneloop}
{L. Br\"ucher, J. Franzkowski and D. Kreimer}.
\newblock {\em Mod. Phys. Lett.} {\bf A9} (1994) 2335--2346;
{D. Kreimer}.
\newblock {\em {Int. J. Mod. Phys.}} {\bf A8} (1993) 1797--1814;
{L. Br\"ucher and J. Franzkowski}.
{\em Mod. Phys. Lett.} {\bf A14} (1999) 881;
{L. Br\"ucher, J. Franzkowski and D. Kreimer}.
\newblock {\em Comp. Phys. Comm.} {\bf 85} (1995) 153--165;
{L. Br\"ucher, J. Franzkowski, D. Kreimer}.
\newblock {\em Comp. Phys. Comm.} {\bf 107} (1997) 281--291.

\bibitem{Zerw2}
{A. Djouadi, J. Kalinowski and P. M. Zerwas}.
\newblock {\em Z. Phys.} {\bf C 70} (1996) 435--448.

\bibitem{rulesInPrep}
{L. Br\"ucher and R. Santos}.
\newblock {\em in preparation}.

\end{thebibliography}



\end{document}